\RequirePackage{rotating}
\documentclass[structabstract]{aa}
\pdfoutput=1

\usepackage{savesym}
\usepackage{caption}
\usepackage{amsmath}
\savesymbol{iint}
\usepackage{txfonts}
\restoresymbol{TXF}{iint}
\usepackage{natbib}
\usepackage{graphicx}
\usepackage{rotating}
\usepackage{tablefootnote}
\usepackage{threeparttable}
\usepackage{float}
\usepackage{placeins}
\usepackage{array}
\usepackage{xcolor}
\usepackage{hyperref}
\newcommand{\Msun}{$M_{\odot}$}

\newcommand{\Lsun}{$L_{\odot}$}

\newcommand{\Mstar}{$M_{\star}$\ }
\newcommand{\Lstar}{$L_{\star}$\ }
\newcommand{\Rstar}{$R_{\star}$\ }

\newcommand{\Rsun}{$R_{\odot}$}

\newcommand{\Ox}{[\oi]$\lambda$6300}

\newcommand{\Ha}{H$\alpha$}
\newcommand{\Hb}{H$\beta$}

\newcommand{\sii}{\ion{S}{ii}}
\newcommand{\oi}{\ion{O}{i}}
\newcommand{\oii}{\ion{O}{ii}}
\newcommand{\oiii}{\ion{O}{iii}}
\newcommand{\Ni}{\ion{N}{i}}
\newcommand{\nii}{\ion{N}{ii}}
\newcommand{\Hi}{\ion{H}{i}}
\newcommand{\heii}{\ion{He}{ii}}
\newcommand{\hei}{\ion{He}{i}}

\newcommand{\neiii}{\ion{Ne}{iii}}

\newcommand{\feii}{\ion{Fe}{ii}}
\newcommand{\caii}{\ion{Ca}{ii}}

\newcommand{\SII}{[\sii]$\lambda\lambda$6716,6731}

\newcommand{\OIII}{[\oiii]$\lambda$5007}

\newcommand{\siib}{[\sii]$\lambda$6731}
\newcommand{\siirat}{[\sii]$\lambda\lambda$ 6716/6731}
\newcommand{\ferat}{[\feii]$\lambda\lambda$ 7155/8617}

\newcommand{\kms}{km s$^{-1}$}
\newcommand{\ndeg}{$^{\circ}$}
\newcommand{\bdec}{H$\alpha$/H$\beta$}
\newcommand{\av}{A$_{V}$}
\newcommand{\eden}{$n_{\mathrm{e}}$}
\newcommand{\xe}{$x_{\mathrm{e}}$}
\newcommand{\Te}{$T_{\mathrm{e}}$}
\newcommand{\Lacc}{$L_{\mathrm{acc}}$}
\newcommand{\Macc}{$\dot{M}_{\mathrm{acc}}$}
\newcommand{\Mout}{$\dot{M}_{\mathrm{out}}$}
\newcommand{\peryr}{yr$^{-1}$}
\newcommand{\nh}{$n_{\mathrm{H}}$}

\newcommand{\percm}{cm$^{-3}$}

\newcommand{\hhw}{HHW$\mathrm{_{2}}$}
\usepackage{natbib}
\graphicspath{{./}}
\usepackage{graphicx,caption}
\usepackage{makecell}

\begin{document}
	\journalname{Astronomy $\&$ Astrophysics}
    \title{Investigating the asymmetry of young stellar outflows: Combined MUSE-X-shooter study of the Th~28 jet \thanks{Based on observations collected with MUSE and X-shooter at the Very Large Telescope on Cerro Paranal (Chile), operated by the European Southern Observatory (ESO). Program IDs: 60.A-9322(A) and 095.C-0134(A)}}

        \author{
                A. \,Murphy \inst{1}
                \and 
                E. T. \,Whelan \inst{2}
                \and 
                F. \,Bacciotti \inst{3}
            \and 
                D. \,Coffey \inst{4}
                \and 
                F. \,Comer\'on \inst{5}
                \and 
                J. \,Eisl\"offel \inst{6}
            \and 
                B. \,Nisini \inst{7}
            \and 
                S. \,Antoniucci \inst{7}
            \and 
                J. M. \,Alcal\'a \inst{8}
            \and 
                T. P. \,Ray \inst{9}}
            
        \institute{Academia Sinica Institute of Astronomy and Astrophysics, No 1. Sec. 4, Roosevelt Rd., Taipei 10617, Taiwan
        \and
        Department of Experimental Physics, Maynooth University, Maynooth, Co Kildare, Ireland
        \and
        INAF—Osservatorio Astrofisico di Arcetri, Largo E. Fermi 5, I-50125 Firenze, Italy
        \and
        School of Physics, University College Dublin, Belfield, Dublin 4, Ireland
        \and
        ESO, Karl-Schwarzschild-Strasse 2, 85748 Garching bei M\"unchen, Germany
        \and
        Th\"uringer Landessternwarte, Sternwarte 5, 07778 Tautenburg, Germany
        \and
        INAF—Osservatorio Astronomico di Roma, Via di Frascati 33, 00078 Monte Porzio Catone, Italy
        \and
        INAF—Osservatorio Astronomico di Capodimonte, via Moiariello 16, 80131 Napoli, Italy
        \and
        Dublin Institute for Advanced Studies, 31 Fitzwilliam Place, D02 XF86, Dublin, Ireland}
        
        \titlerunning{MUSE-X-shooter study of asymmetry in the Th 28 jet} 
        \date{Received 20 June 2024 / Accepted 23 August 2024}
        
        \abstract {Characterising stellar jet asymmetries is key to setting robust constraints on jet launching models and improving our understanding of the underlying mechanisms behind jet launching. }{We aim to characterise the asymmetric properties of the bipolar jet coming from the Classical T Tauri Star Th 28.}{We combined data from integral field spectroscopy with VLT/MUSE and high-resolution spectra from VLT/X-shooter to map the optical emission line ratios in both jet lobes. We carried out a diagnostic analysis of these ratios to compare the density, electron temperature, and ionisation fraction within both lobes. The mass accretion rate was derived from the emission lines at the source and compared with the mass outflow rate derived for both lobes, using the estimated densities and measured \Ox\ and \siib\ luminosities.}{The blue-shifted jet exhibits a significantly higher electron temperature and moderately higher ionisation fraction than the red-shifted jet. In contrast to previous studies, we also estimated higher densities, denoted as \nh, in the blue-shifted jet by a factor of $\sim$2. These asymmetries are traced to within 1\arcsec\ (160 au) of the source in the line ratio maps. We find \Macc\ = 2.4 $\times$ 10$^{-7}$ \Msun\ \peryr, with an estimated obscuration factor of $\sim$54 due to grey scattering around the star. Estimated values of \Mout\ range between 0.66 – 13.7 $\times~ 10^{-9}$ \Msun~\peryr\ in the blue-shifted jet and 5-9 $\times~ 10^{-9}$ \Msun~\peryr\ in the red-shifted jet.}{The emission line maps and diagnostic results suggest that the jet asymmetries originate close to the source and are likely to be intrinsic to the jet. Furthermore, the combined dataset offers access to a broad array of accretion tracers. In turn, this enables a more accurate estimation of the mass accretion rate, revealing a value of \Macc\ that is higher by a factor $>$ 350 than would otherwise be determined.} 
 
        \keywords{ISM: jets and outflows -- stars: pre-main-sequence -- stars: individual: Th~28, Sz102}
    
    \maketitle
        \section{Introduction}

    Protostellar jets and outflows often exhibit marked asymmetries between their two lobes, in terms of brightness, morphology, velocity, density, and excitation \citep{Hirth1994}. These observations favour magnetohydrodynamic (MHD) models of jet launching, which have reproduced asymmetric jet launching as a result of the magnetic field configuration across the disc \citep{Matsakos2012, Bai2017, Bethune2017}. Understanding jet asymmetries and incorporating these into jet launching models are  key to building our understanding of the underlying mechanisms of jet launching and the jet-disc connection. 
    
    To provide constraints for these models, we require observations close to the jet source to distinguish the intrinsic properties of the jet from the effects of environmental interaction. In particular, optical and near-infrared (NIR) forbidden emission lines (FELs) as well as the permitted H lines trace the high-velocity jet in the inner outflow. Examining the flux ratios between these lines makes it possible to measure a range of properties, such as the gas density, temperature, and ionisation fraction, with different line ratios providing measures sensitive to regions with higher or lower density, temperature, and so on. Some FELs of [\sii] and [\oi] have also shown a low-velocity emission component (LVC), which itself comprises both broad and narrow components (BC and NC, respectively). The BC is thought to trace a disc wind, while the NC can be associated with disc emission or with the disc wind \citep{Banzatti2019, Nisini2024}; in the latter case, it has even been shown to have the same position angle (PA) as the high-velocity jet \citep{Whelan2021}.
    
    Emission line diagnostics allow us to constrain the conditions in the launching region, the shock properties of the jets, and their mass outflow rates, which are key to understanding both the jet launching mechanism and the jet asymmetries. In particular, spectral studies have previously highlighted the value of combined multi-wavelength diagnostics in order to access emission lines and line ratios tracing different regions of the jet \citep{Nisini2005, Podio2006, Bacciotti2011, Giannini2019}. Furthermore, by combining data with good spatial coverage from integral field spectroscopy (IFS) and high-resolution broadband spectra, we can disentangle different spatial and kinematic components; for example: resolving shock regions, identifying emission components associated with the high-velocity jet (the HVC) and the LVC, and examining how the jet properties vary as it propagates along the outflow axis.
    
    \begin{figure}
        \centering
        \includegraphics[width=9cm, trim= 0cm 0cm 0cm 0cm, clip=true]{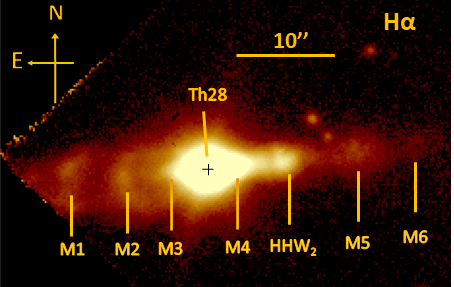}
        \caption{2014 MUSE observation of Th~28, displaying the jet in H$\alpha$ emission before continuum subtraction, adapted from Fig. 1 of \citet{Murphy2021}. The positions of the source (black cross) and all knots discussed in this paper are indicated.}
        
        \label{fig:halpha_snapshot}     
        \end{figure}
     
    However, characterising both jet lobes close to the source typically poses observational challenges. For example, one side of an asymmetric jet may be substantially fainter or more obscured by the surrounding envelope. As a result, models of jet-disc interactions that attempt to reproduce these features rely on constraints from a handful of examples, where it may be unclear whether the observed asymmetries are due to environmental or intrinsic factors \citep[e.g., the relatively well-studied RW Aur jet; see][]{Woitas2002, Melnikov2009}.   
    
	The goal of this paper is to combine observations from the Multi-Unit Spectroscopic Explorer (MUSE) and X-shooter \citep{Bacon2010, Vernet2011}, mounted on UT4 and UT3 of the Very Large Telescope (VLT), respectively, to obtain a multi-wavelength view of the asymmetric bipolar jet from the Classical T Tauri Star (CTTS) Th 28 \citep{Krautter1986}. This jet is of special interest for several reasons: 1), the strong asymmetries in the properties of the two jet lobes, which may be connected to the launching region conditions on either side of the disc; 2) previous detections of a possible rotation in the jet, which is a key prediction of MHD jet launching models \citep{Coffey2004}; 3) the small-scale precession detected by \citet{Murphy2021}, which may be linked to the presence of a brown dwarf companion in the inner disc; and 4) the apparent transverse asymmetry in electron temperature and density detected across the jet axis by \citet{Coffey2008}. The low inclination of the jet presents challenges in differentiating  kinematic structures, but makes it feasible to study both lobes close to the launching region, due to the edge-on nature of the disc, thus offering the possibility to disentangle the cause of the observed asymmetries. For example, we can consider whether asymmetric shocks are able to produce the apparent precession and rotation signatures.  However, to this end, we need to be able to compare not only the previously observed morphology of each lobe and their detailed kinematics, but also identify emission line tracers that enable us to study their density, excitation, and shock properties. This is a particular challenge in the blue-shifted lobe, where most emission lines are observed very faintly if at all; therefore, it is especially important to identify emission lines which trace as much of this jet as possible.
    
    This paper follows \citet{Murphy2021}, who presented morphological and kinematic results from MUSE spectro-imaging of Th 28. For this diagnostic study, we complement the emission lines available in the MUSE data with X-shooter spectra. These added data will allow us to: 1) access a broader array of diagnostic line ratios sensitive to regions with different density and excitation conditions in the jet; 2) use the enhanced velocity resolution of the X-shooter observations to search for distinct velocity components in this low-inclination jet; and 3) simultaneously access a broader array of emission lines tracing both the jet and accretion. On the other hand, the MUSE spectro-images allow us to examine the variation in conditions between different spatial regions of the jet, while the combined datasets enable us to obtain position-velocity (PV) maps both along and transverse to the jet axis. These can be used to investigate how the jet properties vary both with respect to the velocity (i.e., in different kinematic components) and spatially, either along the direction of propagation or across the jet axis. Combining the MUSE data of Th 28 with X-shooter observations at high spectral resolution is therefore an optimal strategy. This approach allows us to combine the detailed spatial information provided by MUSE with the kinematic detail provided by X-shooter, enabling us to gain a more complete three-dimensional picture of both jet lobes and the differences between them. 
        
        \section{Target, observations, and data}
        
        \subsection{Th~28}
        \label{subsection:th28_background}

        Th 28  (\object{ThA 15-28}; also referred to as Sz 102 or Krautter's Star) is located in the Lupus III cloud (D $\sim$ 160 pc \citep{Dzib2018}).  The key parameters of the star are summarised in Table 1 of \citet{Murphy2021}. It has a nearly edge-on disc and drives a bright bipolar jet (HH 228) in the plane of the sky \citep{Graham1988, Louvet2016}.  Due to the disc orientation, the star appears to be very under-luminous at 0.03 L$_{\odot}$ \citep{Mortier2011} and the stellar spectrum is heavily veiled with weak absorption lines, indicative of significant mass accretion activity.
        
    The Th~28 bipolar jet lies in the east-west direction with PA $=$ +95$^{\circ}$ for the blue-shifted lobe and -85\ndeg\ for the red-shifted lobe \citep{Graham1988}. The jet and counter-jet are noticeably asymmetric in multiple respects, including the morphology, kinematics, and gas properties. A more complete discussion of the morphological and kinematic differences between the two jet lobes is given in \citet{Murphy2021}. In brief, the blue-shifted lobe shows both higher radial and tangential velocities (measured through knot proper motions) than the red-shifted lobe, with an estimated true jet velocity of $\sim$360 \kms\ versus 270 \kms\ in the red-shifted lobe. This difference is more pronounced in the radial velocities (which differ by a factor of $\sim$4) than in the tangential velocities estimated from proper motions. We also note that close to the source, there appears to be significant scattering, particularly of emission from the bright red-shifted jet which results in contamination of the blue-shifted jet spectra.

    Diagnostic studies overall suggest a less dense, hotter and more ionised blue-shifted jet compared with the denser, cooler and less ionised red-shifted jet. Estimates of the electron density based on emission lines of [\nii], [\sii] and [\oi] show higher values (6.7 $\pm$ 2.2 $\times$ 10$^{4}$ \percm\ to $\geq$ 1 $\times$ 10$^{5}$ \percm) in the blue-shifted jet compared to those found in the red-shifted jet (1 $\times$ 10$^{4}$ \percm\ to 4.6 $\pm$ 2.4 $\times$ 10$^{4}$ \percm) \citep{Liu2014,Liu2021}. On the other hand, measurements by \citet{Coffey2010} using [\feii] lines show 1.4 $\pm$ 0.4 $\times$ 10$^{4}$ \percm\ and \eden\ = 2 $\pm$ 0.1 $\times$ 10$^{4}$ \percm\ in the blue- and red-shifted jets, respectively. 
   
     Measurements of the ionisation and electron temperature for Th 28 are more limited, but tend to suggest \Te\ $\sim$ 1.5 $\times$ 10$^{4}$ K to 2 $\times$ 10$^{4}$ K in both lobes, with the blue jet slightly hotter perhaps \citep{BE99, Coffey2008, Liu2014}. For example, \citet{Liu2014}  measure \Te\ of 1.7 $\pm$ 0.6 $\times$ 10$^{4}$ K in the blue-shifted jet and 1.5 $\pm$ 0.5 $\times$ 10$^{4}$ K in the red-shifted jet. These values are again similar to the results found for high-velocity jets by \citet{Giannini2019}. The optical line ratios obtained by \citet{Liu2014} also suggest the  ionisation fraction of the blue-shifted jet is higher by a factor of 2-3. Consequently, despite the higher values of \eden\ in the blue-shifted jet, they estimate the total density to be a factor 1.4-2 higher in the red-shifted jet. \citet{Liu2014} trace these asymmetries to within 50 au of the source, which suggests they are intrinsic to the jet launching. 
  
    One limitation in comparing previous studies of the Th 28 jet is that they typically measure the jet properties at different distances from the source and often in only the brighter red-shifted lobe. Due to the narrow width of these jets (typically 0\farcs15-2.0\arcsec) they are also notoriously difficult to resolve transversely. Nevertheless, transverse maps of \eden\ \citep{Coffey2008} also show some indications of spatial asymmetries across the jet cross-section. These results show the [\sii] ratio saturating on one side of the red-shifted jet axis (indicating a lower \eden\ limit of 2.5 $\times~10^{4}$ \percm) while it falls close to the low density limit elsewhere (\eden\ $\sim$ 50 \percm).
    
    \subsection{Observations}

    We have made use of VLT/MUSE and X-shooter observations of the Th~28 jet obtained in 2014 and 2015, respectively. The details of both datasets are summarised in Table \ref{table:observations}. The MUSE data comprise wide-field mode observations obtained during the science verification period and they are not adaptive optics (AO) corrected. The seeing-limited resolution is thus 0\farcs9 and the velocity resolution across the MUSE wavelength range varies from 170~km s$^{-1}$ (4750 \AA) to 80~km s$^{-1}$ (9350 \AA). Figure \ref{fig:halpha_snapshot} shows the inner jet region and the positions of the knots detected in the 2014 MUSE data. These data and the reduction are described in further detail in a previous paper \citep{Murphy2021}.
    
    The X-shooter observations were obtained on the 17 April 2015 (Program ID: 095.C-0134(A)). The echelle slit nodding mode was used with a slit width of 0\farcs4 to 0\farcs5; the slit was aligned along the jet axis to obtain two sets of exposures, one centred on the blue-shifted jet lobe (position A) and the other on the red-shifted jet lobe (position B). The average seeing was 0\farcs9 with a velocity resolution of approximately 15 \kms\ in the optical (VIS) arm and 30 \kms\ in the ultra-violet (UV) and NIR arms. The total integration times for each nodding position on the UV, VIS, and NIR arms were 1200s each.

    \begin{table*}
        \begin{threeparttable}
        \centering
        
        \caption[observations]{Observation details.}
 
        \renewcommand{\arraystretch}{1.4}
        \begin{tabular}{{p{0.11\textwidth}<{\raggedright} p{0.13\textwidth}<{\raggedright} p{0.06\textwidth}<{\raggedright} p{0.08\textwidth}<{\raggedright} p{0.08\textwidth}<{\raggedright} p{0.12\textwidth}<{\raggedright} p{0.05\textwidth}<{\raggedright} p{0.08\textwidth}<{\raggedright} p{0.07\textwidth}<{\raggedright}}}
                \hline \hline
                Date &  Instrument &  PA (\ndeg) & Seeing & Slit width & Spatial sampling & Arm & $\tau_{exp}$ (s) & R \\
        \hline
                  2014-Jun-23  & VLT/MUSE  & 45 & 0\farcs9 & ... & 0\farcs2 & ... & 3600 & 2680 \\
        2015-Apr-2015   & VLT/X-shooter   & -95 & 0\farcs9 - 1\arcsec\ & 0\farcs5 & 0\farcs16 & UV  & 2 $\times$ 1200$^{1}$ & 9100 \\ 
        ...              &                 & ...  &  ...   & 0\farcs4 & 0\farcs16   & VIS & 2 $\times$ 1200$^{1}$ & 17400 \\ 
        ...              &                 & ...  & ...   & 0\farcs4 & 0\farcs2             & NIR & 2 $\times$ 1200$^{1}$ & 10500 \\ 
                \hline
        \end{tabular} 
        \label{table:observations}
        \begin{tablenotes}
                \item[] \textbf{Notes:} (1) Total exposure times for the X-shooter observations include both nodding mode positions.
                \end{tablenotes}
            \end{threeparttable}
    \end{table*}
    
    \subsection{Data reduction and analysis}
        \label{section:reduction}
        
    The reduction and calibration of the MUSE data are described in the preceding paper; however, we note that the dataset comprises one complete datacube covering the full wavelength range (full datacube) and a set of cubes produced using an updated version of the pipeline, which excludes several emission lines at wavelengths $<$ 5000 \AA\ and $>$ 8500 \AA\ (the partial datacubes). Our analysis (both here and in the preceding paper) primarily utilises the partial datacubes. However, we have made use of the original datacube to access additional emission lines such as \Hb.

    The X-shooter observations were reduced using v. 2.3.0 of the ESO X-shooter reflex pipeline, including flux and wavelength calibrations, following the procedure described in \citep{Alcala2017}.  As part of this procedure, short exposures with the 5\farcs0 slit were used to correct for slit losses and obtain accurate spectrophotometry. The data were reduced in stare mode to retain two sets of spectra centred on the two jet lobes.
    
    We carried out an additional check on the wavelength calibration of the UV and VIS observations, making use of bright sky emission lines identified along the full wavelength range of the spectra. We sampled these lines in regions containing no jet emission and fit them with a Gaussian profile to estimate the velocity calibration error. In the VIS arm, we found consistent offsets of -6 to -8 \kms\ in the sky lines, with median values of -7.5 \kms\ and 5.7 \kms\ for the A and B positions, respectively. These correspond to approximately one pixel in the wavelength axis; therefore, we corrected our spectra for these median offsets. In the UV arm, only the bright [\oi]$\lambda$5577 line had a sufficient signal-to-noise (S/N) to estimate a velocity offset, which gives an average correction of -1.6 \kms. In addition, we applied a barycentric velocity correction of +18.9 \kms\ to the X-shooter spectra, as well as correcting for the LSR velocity of +4 \kms\ \citep{Liu2021}. We quote all velocities hereafter with respect to the LSR frame.

    To remove the sky emission, we median-averaged the spectrum over spatial regions with minimal jet emission in each observation, masking portions of the spectrum at selected wavelengths where the jet emission extends along the full spatial axis (i.e., the \Ha\ and optical [\sii] emission lines). This averaged spectrum was then subtracted from all spatial positions. The continuum subtraction was carried out by fitting a third-order polynomial to regions of continuum-only emission in the on-source spectrum. The fitted continuum at each spatial position was then subtracted from the spectrum.
    
    \section{Emission lines in the Th 28 bipolar jet}
    
    \subsection{Detected emission lines}
    The MUSE observations show the Th 28 jet between 4800 \AA\ and 9000 \AA. The observed jet morphology is described in \citet{Murphy2021}; here, we briefly summarise the results, while the inner jet region relevant to this study is shown in Fig. \ref{fig:halpha_snapshot}. In general, the brightest inner regions of both jets are observed in FEL and \Hi\ emission up to 15\arcsec\ ($\sim$2400 au) from the source, with the blue-shifted lobe being significantly fainter and characterised by bow shocks visible in lines of \Hi, [\sii], [\nii] and [\oiii], located at separations of -3\farcs3, -8\farcs8 and -14\farcs2 from the source position. The red-shifted jet is brighter, more collimated and shows more compact knots in a wider array of FELs.
    
    In the X-shooter data, both jet lobes are detected to approximately 3-4\arcsec\ in both the UV and VIS arms. A full catalogue of detected jet emission lines as well as the corresponding position-velocity (PV) maps are included with the supplementary material\footnote[1]{\url{https://doi.org/10.5281/zenodo.13373809}}. These maps show the same velocity asymmetry between the jet lobes which is detected in previous observations, including our previous measurements using the MUSE data. They consistently show that the red-shifted jet has radial velocities of $\sim$ +20-30 \kms\ whereas the blue-shifted lobe has radial velocities that are a factor of $\sim$4 times higher at -80-100 \kms. These maps also illustrate that the full width at half maximum (FWHM) of the emission in both lobes (approximately 120 \kms\ in the inner red-shifted jet and 200 \kms\ in the blue-shifted jet) combined with the low inclination angle (and, hence, lower jet radial velocities) cause each of them to be visible in both the red- and blue-shifted velocity channels, as observed in \citet{Murphy2021}.

    The blue-shifted jet is again significantly fainter in most emission lines. In the UV arm, both lobes show emission from species such as \caii, [\oii], [\oiii], [\sii], and \Hi, and both also show [\neiii] emission within about 1\farcs5 of the source consistent with the detection by \citet{Liu2014}. While the MUSE observations showed [\oiii]$\lambda$5007 emission from the blue-shifted lobe only, the X-shooter spectra make it clear that there is red-shifted [\oiii] emission ($v_{\mathrm{rad}}$ = +17 \kms), which appears to be from the inner region of the red-shifted jet. The [\Ni]$\lambda$5200 doublet is also detected extended along the red-shifted jet axis, however in our data it appears to be blended with another emission line. The Ca H and K lines are similarly detected in this lobe; the H line is concentrated near the jet base and is blended with the H$\epsilon$ line, while the K line is clearly extended along the outflow axis.
    
    In the VIS arm, the jet is traced by bright emission lines of \Hi, [\sii], [\oi], and [\nii].  We confirm the presence of emission from refractory species such as [\ion{Ni}{ii}], [\feii], and [\caii] near the base of the red-shifted jet, with a few of these lines (e.g., [\feii]$\lambda\lambda$8716, 7155 and [\caii]$\lambda$7292) extending as far as 3\arcsec. As in the MUSE data, these species are only clearly identified in the red-shifted lobe. A few of the brightest optical emission lines \Ha, [\nii]$\lambda$6583, and [\sii]$\lambda\lambda$6716, 6731 extend to the limit of the slit (8\arcsec\ from the source position) in the red-shifted lobe. The [\oi]$\lambda$5577 line is also present close to the source and shows a low-velocity blue-shifted emission peak slightly offset along the blue-shifted jet axis. Between both arms, we detected numerous low-excitation emission lines overlapping with those detected in the MUSE data, enabling us to carry out complementary diagnostics with both datasets.
    
        \subsection{[\oi] and [\oiii] emission from the jet}
    \label{subsection:oi_emission}

    \begin{figure*}
    \centering
    \includegraphics[width=15cm, trim= 0.4cm 0.4cm 0cm 0cm, clip=true]{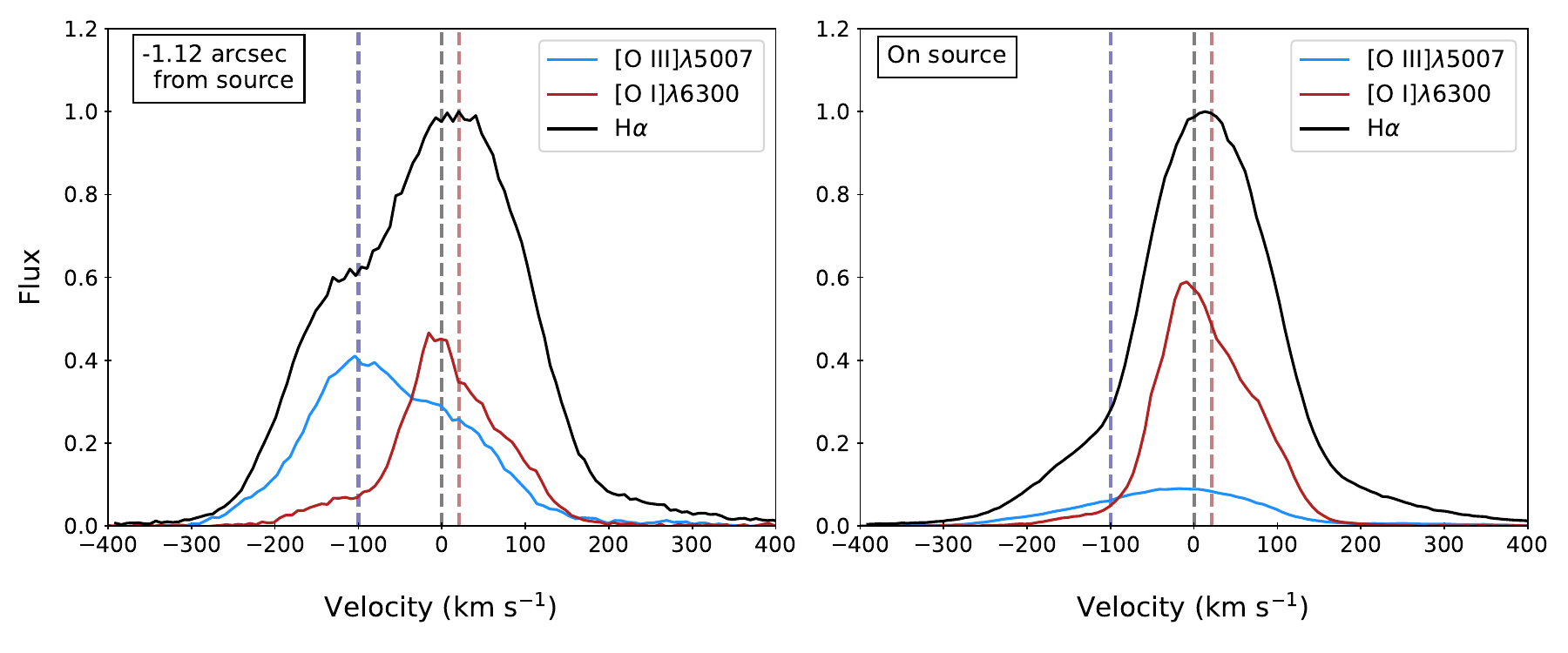}
        \caption{Emission line profiles at two spatial positions in the X-shooter data. Left:\  Blue-shifted jet; right, the source position. All velocities are given with respect to the LSR. The estimated radial velocities of the blue- and red-shifted lobes are shown with vertical blue and red dashed lines at -100 \kms\ and +20 \kms, respectively, while the 0 \kms\ velocity position is shown with the vertical black line.}
    
        \label{fig:jet_lineprofiles}     
        \end{figure*}
    
    \begin{figure*}
    \centering
        \includegraphics[width=18cm, trim= 0cm 0cm 0cm 0cm, clip=true]{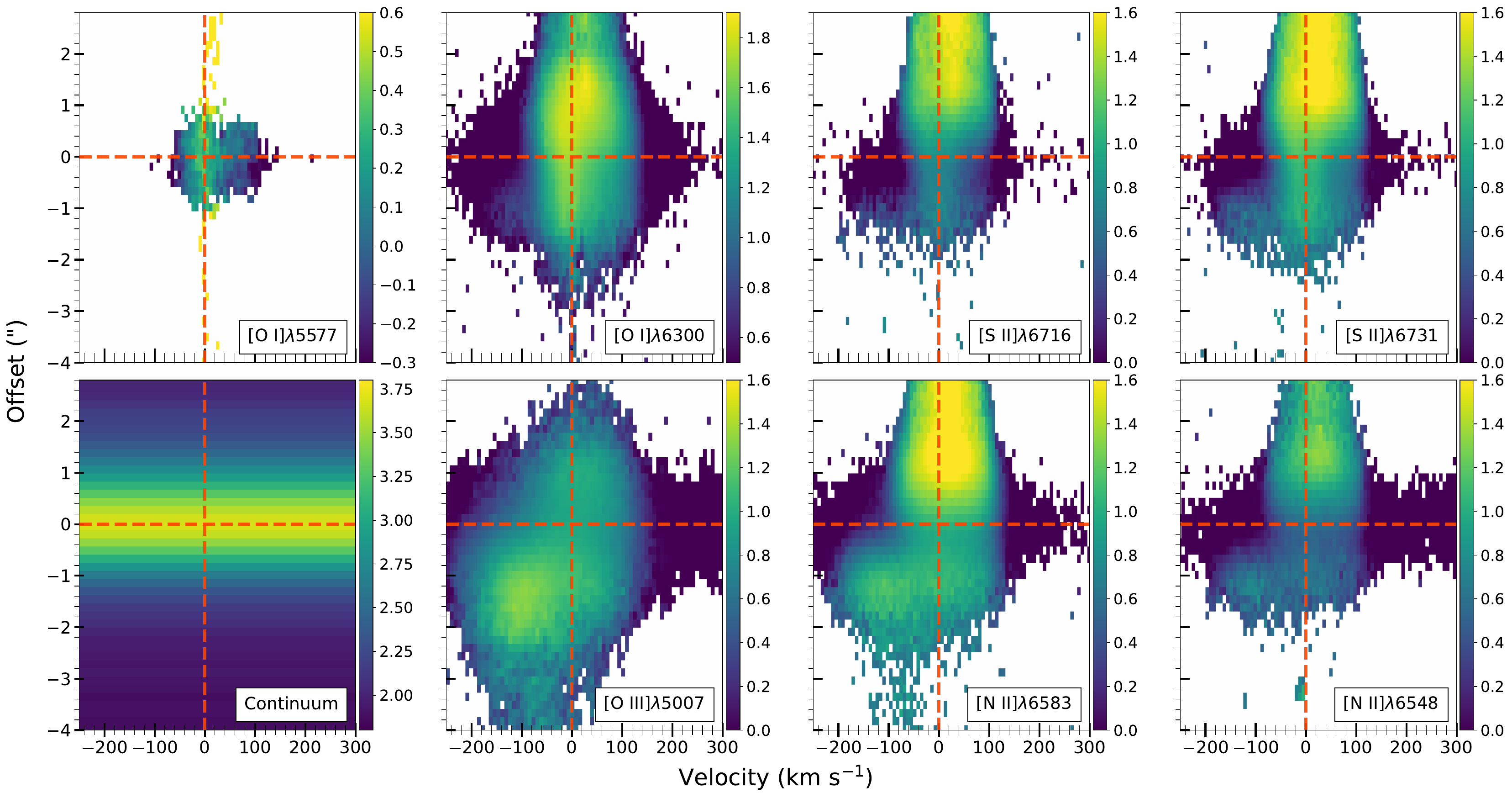}
        \caption{PV maps of the line-to-continuum emission ratio in eight FELs, obtained from the X-shooter data. All images are shown in log scale and aligned with the blue-shifted lobe toward the bottom of the panel (negative-y direction). Top: [\oi] and [\sii] lines; [\oi]$\lambda$5577 primarily traces the LVC while [\oi]$\lambda$6300 and [\sii] trace both the high-velocity jet and LVC. Bottom: Far-left panel shows an image of the continuum emission flux. The remaining panels show [\oiii] and [\nii] lines that trace the high-velocity jet only.}
        \label{fig:scattering_maps}     
        \end{figure*}

    \begin{figure*}
    \centering
        \includegraphics[width=9cm, trim= 0cm 0cm 0cm 0cm, clip=true]{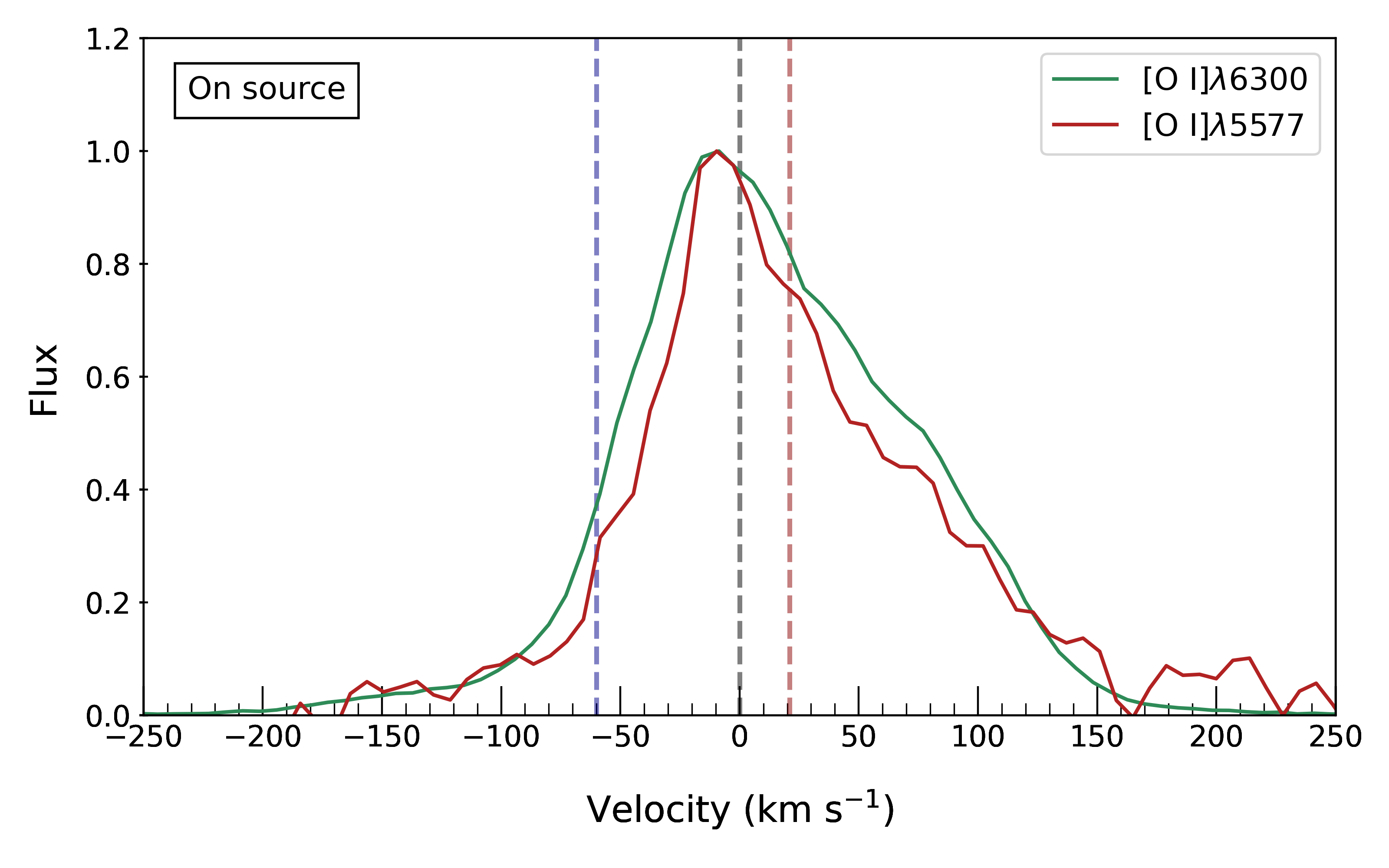}
        \includegraphics[width=6.4cm, trim= 0cm 0cm 0cm 0.1cm, clip=true]{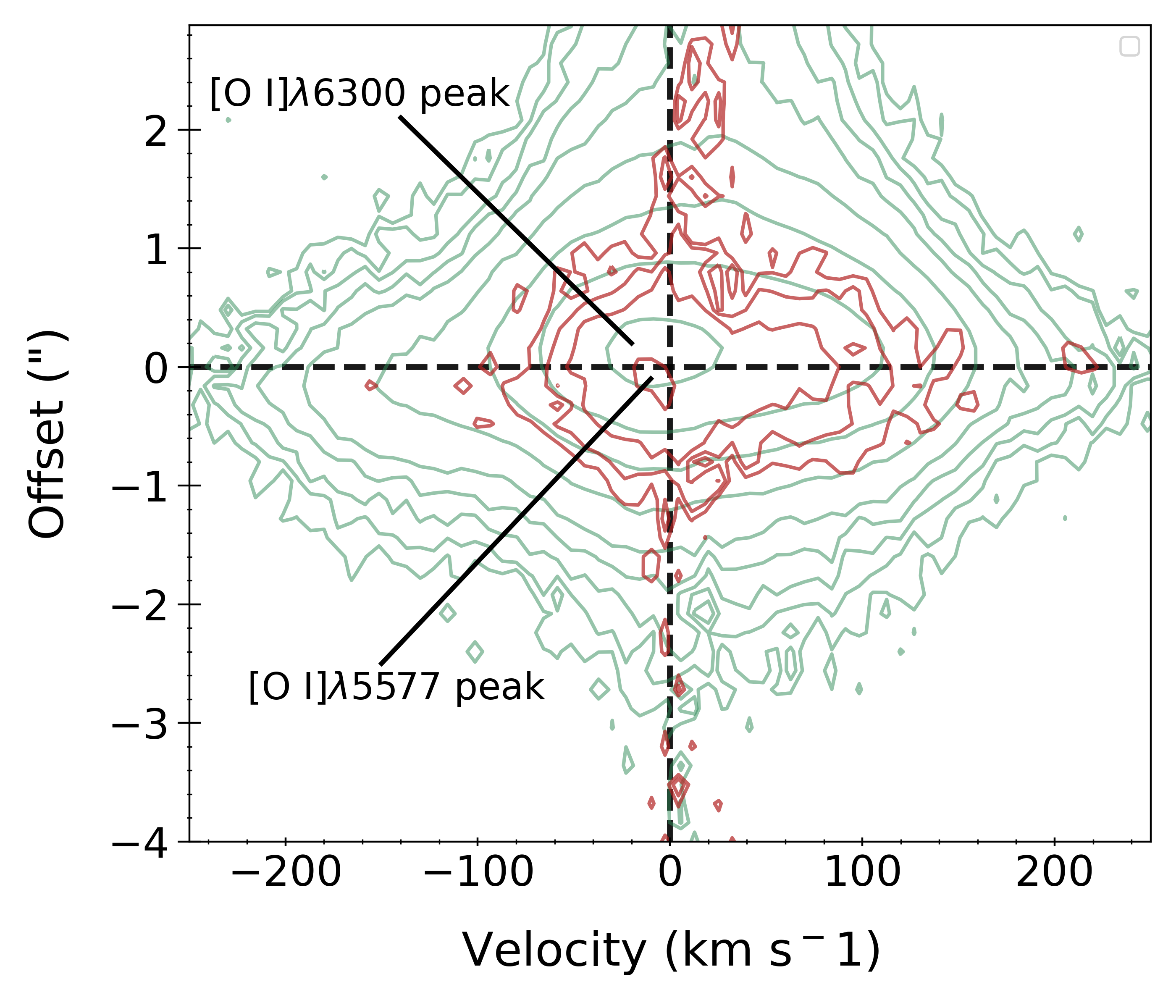}
    \caption{Comparison of the [\oi]$\lambda$6300 and [\oi]$\lambda$5577 emission in the X-shooter data. In both panels [\oi]$\lambda$6300 is shown in green, [\oi]$\lambda$5577 in red. Left:  Flux-normalised line profiles sampled at the source position.. The estimated radial velocities of the blue- and red-shifted lobes are shown with vertical blue and red dashed lines at -100 \kms\ and +20 \kms, respectively, while the 0 \kms\ velocity position is shown with the vertical black line. Right: PV maps from the X-shooter spectra showing the blue-shifted jet lobe, with contours starting at 3$\sigma$ of the background level. On both axes the 0 position is marked with a black dashed line.}
        \label{fig:o_profiles}     
        \end{figure*}

    Our diagnostic study makes use of emission line flux ratios to constrain the density and excitation within the jet. In particular, we sought to use the diagnostic method described by \citet{BE99}, which requires the \Ox\ flux as a key input and is not valid in regions where [\oiii] is detected. We also wished to check for contamination by the brighter red-shifted jet. It is therefore important to establish where each of these emission lines is detected in the jet. The low inclination angle of the Th 28 jet means that the velocity shift between different contributions to the emission (e.g., from the two different jet lobes) is also low compared with the MUSE velocity resolution of $> 100$ \kms. 
    
    Therefore, we used the X-shooter spectra with spectral resolution $\sim$ 15 \kms\ to distinguish the emission components. We examined the line profiles of both [\oi] and [\oiii] obtained from the X-shooter spectra, sampled at the source position and the inner regions of each jet lobe. These are shown in Fig. \ref{fig:jet_lineprofiles}, over-plotted with \Ha\ which  highlights broad bow shock structures, such as those seen in the blue-shifted jet morphology. We also examined several jet-tracing FELs using line-to continuum ratio maps in order to highlight the flux from the jet over the scattered emission close to the source position. These are shown in Fig. \ref{fig:scattering_maps}.
We observed two key features from these figures, described below. 

First, extended [\oiii] emission is observed in both the inner red-shifted jet and the blue-shifted jet. In the red-shifted lobe this line is centred at the jet radial velocity and extends to approximately 2\arcsec. In the blue-shifted lobe, this line exhibits a broad peak around -1\arcsec\ from the source. This coincides with a knot detected by \citet{Melnikov2023} and suggests that this is a fast shock that is unresolved in the MUSE data. In Fig. \ref{fig:jet_lineprofiles} we note that the [\oiii] emission shows a broad double-peaked profile, with the highest peak also at $\sim$ -100 \kms\ and a low-velocity (LV) peak extending into high red-shifted velocities. From a comparison with the bow shock models of \citet{Hartigan1987}, we find that both this double-peaked profile and the corresponding H$\alpha$ profile in this knot are characteristic of emission from a bow shock observed at low inclination to the plane of the sky ($i <$ 30\ndeg). This is consistent with the bow shock morphology and [\oiii] emission seen in the other knots of this jet lobe resolved in the MUSE spectro-images and visible in Fig. \ref{fig:halpha_snapshot}. 

Second, a very strong blue-shifted LV peak is observed in \Ox\ (Fig. \ref{fig:scattering_maps}) which extends from the source position to a distance of -2\arcsec\ in the blue-shifted lobe. A similar feature is seen in the [\sii] lines, but not in the [\nii] lines which also show bright emission from the red-shifted jet. This suggests that this feature is not attributable to contamination from scattered emission from the brighter lobe. Instead it may represent the narrow component (NC) of the LVC seen in other jets. While the broad component is thought to trace an MHD disc wind originating at larger disc radii, the NC of RU Lupi has been shown by \citet{Whelan2021} to align with the jet PA, and this may represent an interaction region between the central jet and the surrounding wind (Birney et al., in prep.). Indeed, a blending of the strong LV peak and the red-shifted HV jet may explain the plateau around rest velocity observed by \citet{Comeron2010} only in [\sii] and [\oi]. Examining Fig. \ref{fig:scattering_maps}, we find that this spatially extended, low-velocity feature is observed only in the lines known to trace the LVC ([\sii] and [\oi]) but not in the higher excitation [\nii] and [\oiii] lines which trace the HVC.
    
    Therefore, we examined the [\oi]$\lambda$5577 line, which is a well-established tracer of the LVC but less often detected in the high-velocity jet due to its high critical density ($\sim$10$^{8}$ \percm). Figure \ref{fig:o_profiles} compares the line profiles of this line and [\oi]$\lambda$6300 in the inner blue-shifted jet (left panel) and their PV maps (right panel). We find very similar profiles between the two lines, with [\oi]$\lambda$5577 also showing a low-velocity blue-shifted peak at $\sim$ -10 \kms\ as well as a red-shifted wing. In the contour plots, we also see that the emission in this line peaks close to the continuum position, but with a small spatial offset along the blue-shifted jet direction, supporting the possibility that this emission is associated with the LVC on this side.
        
    This therefore suggests that the [\oi] emission seen in the blue-shifted jet may be a combination of emission from the bow shock and a relatively bright LVC close to the source. This poses challenges in carrying out diagnostics that make use of the \Ox\ line with respect to ensuring that we are sampling line ratios from the same outflow component. In the slower red-shifted jet, the LVC and HVC also cannot be distinguished, but this is less of a concern as the \Ox\ emission from the jet itself is brighter; whereas in the blue-shifted jet the LV peak is a relatively bright component of the observed emission. For diagnostics using this line in the blue-shifted jet, we therefore seek to sample only emission from the high-velocity peak seen around -100 \kms\ in \Ox\, which is most likely to represent the true jet emission. To do so, we used the MUSE datacubes and sample the emission from velocity channels corresponding to velocities below -30 \kms\ to exclude the low-velocity emission.

    \section{Optical diagnostics}
    \label{section:diagnostics}
    
    The wide array of jet-tracing emission lines in the MUSE and X-shooter data allows us to investigate the electron density and temperature in the jets using multiple line ratios. Utilising these line ratios has the advantage of being direct tracers of the conditions in the emitting gas. However, to estimate the mass outflow through the jet lobes, we also need to estimate the hydrogen density (\nh\ = $n_{\mathrm{e}}/x_{\mathrm{e}}$), and hence we require a method of estimating the ionisation fraction, \xe. To do so, we employed the technique detailed in \citet[][hereafter, "the BE method"]{BE99}, which combines multiple line ratios of O, S, and N to determine \xe\ and, hence, \nh. In Sects. \ref{subsection:density} and \ref{subsection:excitation}, we first present the full set of line ratio maps as the primary results obtained from the data; we then present the results of our BE method analysis in Sect. \ref{subsection:bemethod}. Finally, in Sect. \ref{subsection:tv_maps}, we use the same methods to examine transverse maps of the jet properties in order to search for asymmetries across the jet axis.
    
    Here, we provide an overview of the emission line maps and ratios used. We note that since the emission lines used in several of our ratio maps have significantly different wavelengths ($>$1000 \AA), we applied a dereddening correction for extinction to the resulting spectro-images. The estimated corrections are obtained using the \Ha/\Hb\ decrement estimated for shocked gas in each region of the jet \citep{Hartigan1994}; they are described fully in Appendix \ref{section:extinction}. For ratios involving one or more lines that fall outside the range of the partial datacubes (e.g., the [\feii]$\lambda\lambda$ 7155/8617 ratio), all of the images were extracted from the full MUSE datacube to ensure consistency in the pipeline and flux calibrations.
    
    To examine the electron density, we  used ratios of [\sii] and [\feii]. The main ratio in [\sii] is the well-known \siirat, which is proportional to \eden\ between approximately 50 \percm\ and 2 $\times$ 10$^{4}$ \percm\ (where \Te\ $<$ 2 $\times$ 10$^{4}$ K) and is therefore ideal for tracing lower-density jet regions. In the X-shooter data, we also have access to [\sii]$\lambda\lambda$ 4069/4076 and [\sii]$\lambda\lambda$ 4069/6731, which are primarily sensitive to values of \eden\ between $\sim$10$^{5}$ - 10$^{8}$ \percm. To explore the higher-density regions of the jet, we can also look at the ratio [\feii]$\lambda\lambda$ 7155/8617 (hereafter 'the [\feii] ratio'). This is sensitive to regions of \eden\ $\sim$ 10$^{6}$ - 10$^{7}$ \percm, with the theoretical curves for this relationship calculated by \citet{Bautista2015}.

    Two line ratios that can be correlated with temperature are [\nii]$\lambda$(6548+6583)/5755 (hereafter, "[\nii] ratio") and [\oi]$\lambda$(6300 +6363)/5577. Both of these ratios are inversely correlated with the temperature, with almost identical curves as discussed by  \citet{Osterbrock2006}. However, [\oi]$\lambda$5577 is detected only very close to the source position and does not trace the jet. We therefore only show the results for the [\nii] ratio, which is sensitive to \Te\ between 5000 K and 3.5 $\times~10^{4}$ K, with minimal dependence on \eden\ in low-density jet regions with \eden\ $<$ 10$^{4}$ \percm.

    The [\sii]$\lambda$(6716+6731)/[\oi]$\lambda$(6300+6363) ratio (the [\sii]/[\oi] ratio) is sensitive primarily to electron temperatures $>$ 5000 K and to the ionisation fraction in regions below the [\sii] critical density; at higher densities, it also has a strong dependence on \eden; we therefore show this ratio to highlight ionisation in the lower-density regions of the jet \citep{Hartigan1994, Podio2011} and it is one of the diagnostic ratios used to determine these parameters in the BE method (more in the next section).
    
    To further explore the ionisation along the jet we can first use [\nii]$\lambda$(6548+6583)/[\sii]$\lambda$(6716+6731) (the [\nii]/[\sii] ratio), which is primarily dependent on \xe; however it also has a weak positive dependence on \eden\ between densities of 10$^{4}$ to 10$^{6}$ \percm\, and on \Te\ $> 5000$ K. It therefore traces high-excitation regions, increasing where the gas is hotter, denser and more ionised, namely, in shock regions \citep{Hartigan2007}. Finally, the ratio [\nii]$\lambda$(6548+6583)/[\oi]$\lambda$(6300+6363) (the [\nii]/[\oi] ratio) also traces \xe\ and \Te\ between 6000 K and 2 $\times~10^{4}$ K in low-density regions below 10$^{4}$ \percm, with a weak dependence on \eden\  at higher densities (this ratio is also used in the BE method).

    \subsection{Electron density}
    \label{subsection:density}
    
    \begin{figure}
    \centering
        \includegraphics[width=9cm, trim= 0cm 0cm 0cm 0cm, clip=true]{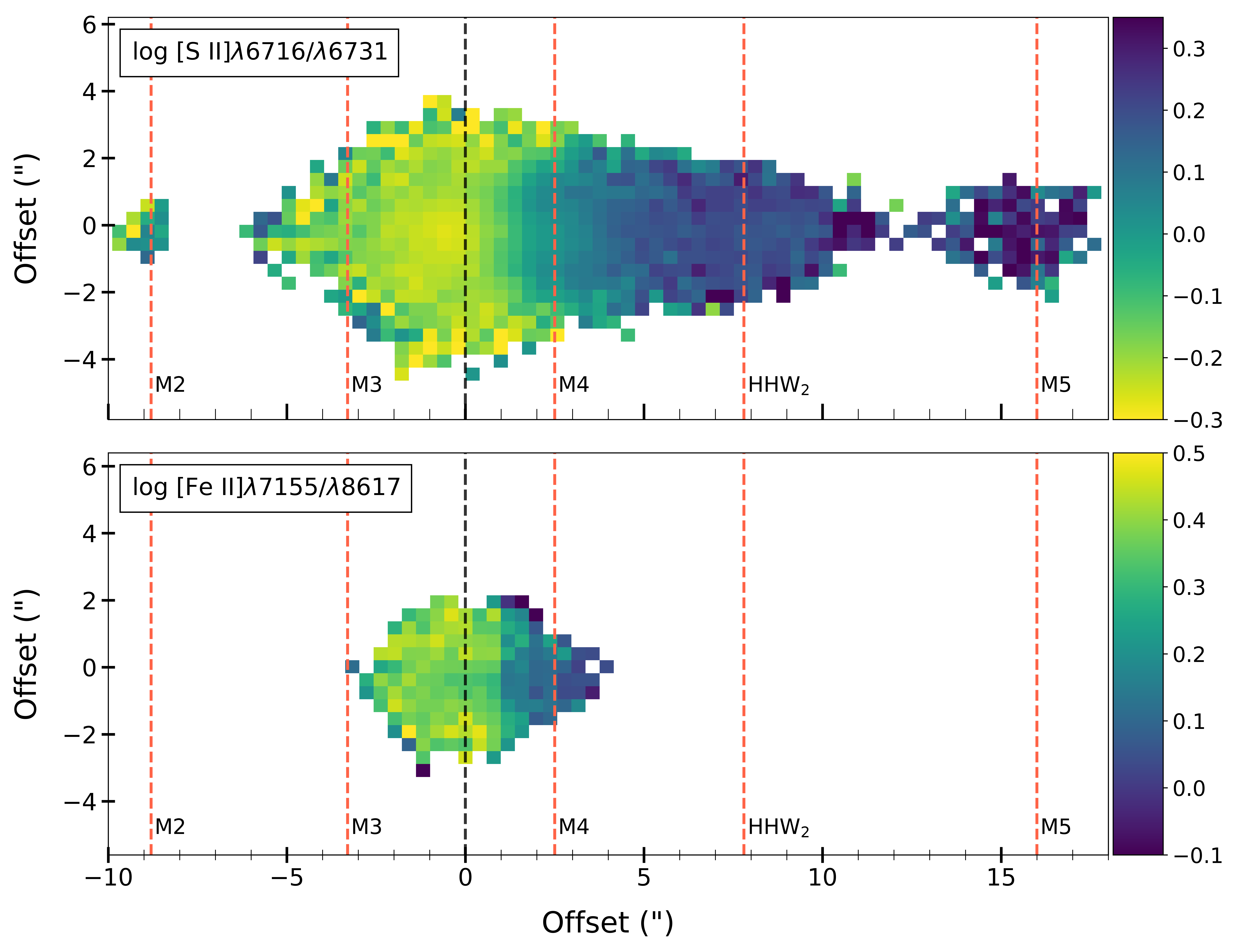}
        \caption{Line ratio maps of the Th 28 jet from the MUSE channel maps (-200 to +140 \kms), showing line ratios which trace \eden. Pixels with flux below 3 $\sigma$ of the background in any of the constituent emission lines have been excluded. Source and knot positions are marked by black and red vertical dashed lines, respectively, with knot positions as defined in \citet{Murphy2021}. Colour scaling is orientated to show the regions of highest \eden\ in yellow. }
        \label{fig:MUSE_density}     
        \end{figure}

\begin{figure}
    \centering
        \includegraphics[width=9.0cm, trim= 1cm 1cm 4.5cm 0cm, clip=true]{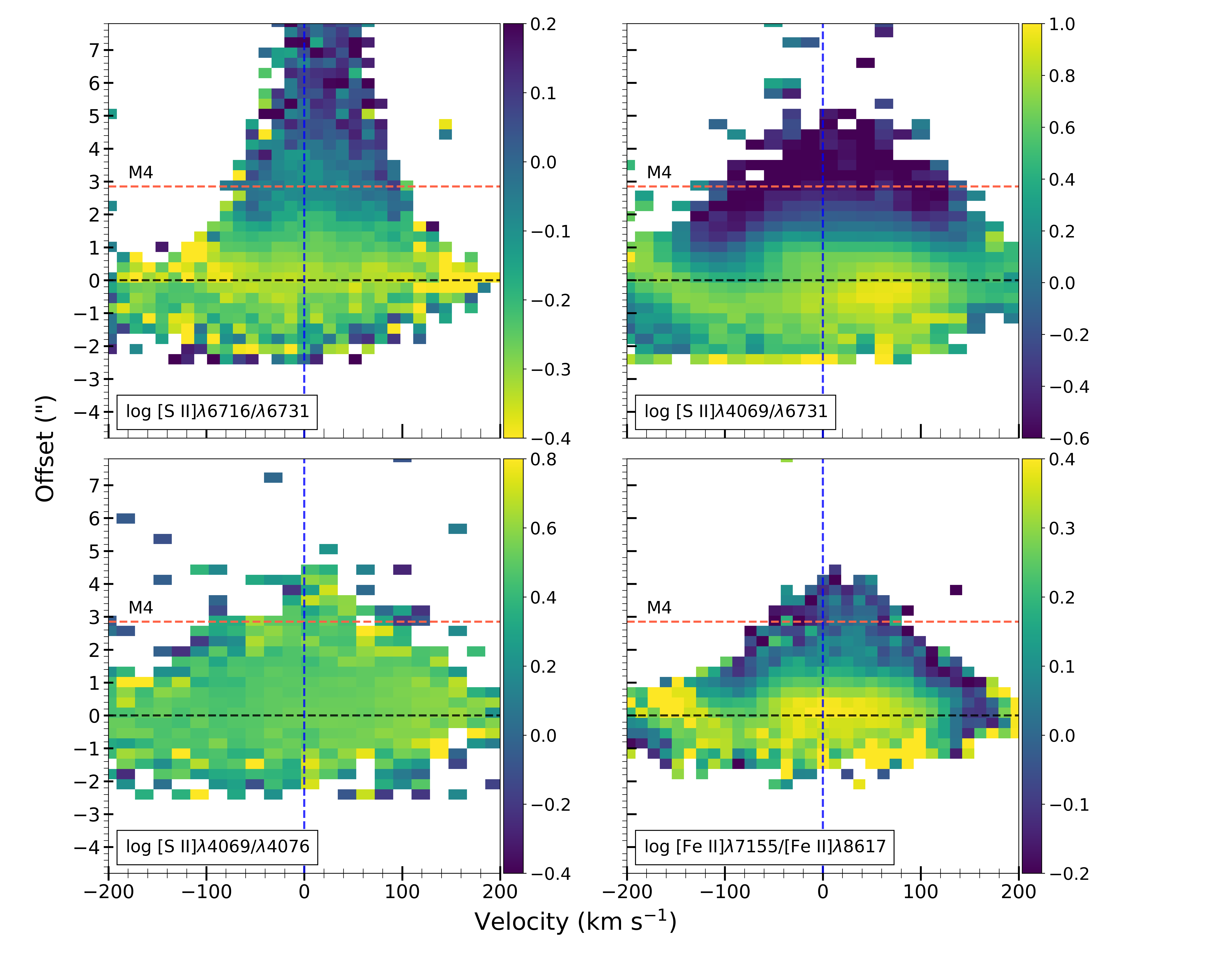}
        \caption{PV maps from the X-shooter data of the Th 28 jet, showing line ratios tracing \eden. The red-shifted jet lobe is located above the source position, with the blue-shifted lobe below. The maps for each line ratio are dereddened and binned along both axes by a factor of 2. Colour maps are as in Fig. \ref{fig:MUSE_density} and the knot M4 is marked with a dashed red line.}
        \label{fig:XS_ne}     
        \end{figure}

    The dereddened ratio channel maps and PV maps are shown in Figs. \ref{fig:MUSE_density} and \ref{fig:XS_ne}, respectively. The ratio channel maps were obtained from the continuum-subtracted MUSE datacubes by summing each of the key emission lines over the combined velocity range of both jet lobes (-200 \kms\ to +140 \kms). These maps cover the full spatial extent of the inner jets, including the bow shock M2 which is clearly detected in [\sii]. The [\feii] ratio is also measured in both microjets up to 2-3\arcsec\ from the source position. Both ratios highlight the jet asymmetry, with the blue-shifted jet lobe showing significantly higher values of electron density \eden. 
    
    The corresponding PV maps are shown in Fig. \ref{fig:XS_ne}. The blue-shifted jet lobe is very faint at separations of $>$ 2\arcsec. In the red-shifted jet, most of the emission lines of interest are too faint to measure flux ratios at more than 4\arcsec\ from the source. One exception is the \siirat\ ratio that is observed across the full extent of the jet. These maps show a general decrease in \eden\ with distance from the source consistent with a drop in jet density. From the blue-shifted channel maps and PV maps, the [\feii] ratio yields values of 1.6-2.5 in the blue-shifted jet, consistent with \eden\ $\sim$ 1 $\times~10^{6}$ \percm; at the source position, this ratio is $>$ 4.0 in the PV maps suggesting \eden\ $\sim$ 1 $\times~10^{7}$ \percm\ \citep{Bautista2015}. In the red-shifted lobe, this ratio drops rapidly from values $\sim 2$ to values below the low-density limit. A similar pattern is seen in the [\sii]$\lambda\lambda$4069/6731 ratio, which again suggests \eden\ $>$ 1 $\times~10^{5}$ \percm\ in the blue-shifted jet and at the source, dropping below the low-density limit $\sim10^{4}$ \percm\ in the red-shifted jet. Finally, in [\sii]$\lambda\lambda$4069/4076, the PV maps show values at or slightly above the low-density limit ($\sim$1 $\times~10^{5.5}$ \percm) both close to the source and in the inner region of both jet lobes.
            
    \subsection{Excitation}
        \label{subsection:excitation}

\begin{figure}
    \centering
        
        \includegraphics[width=9cm, trim= 0cm 0.5cm 0.4cm 0cm, clip=true]{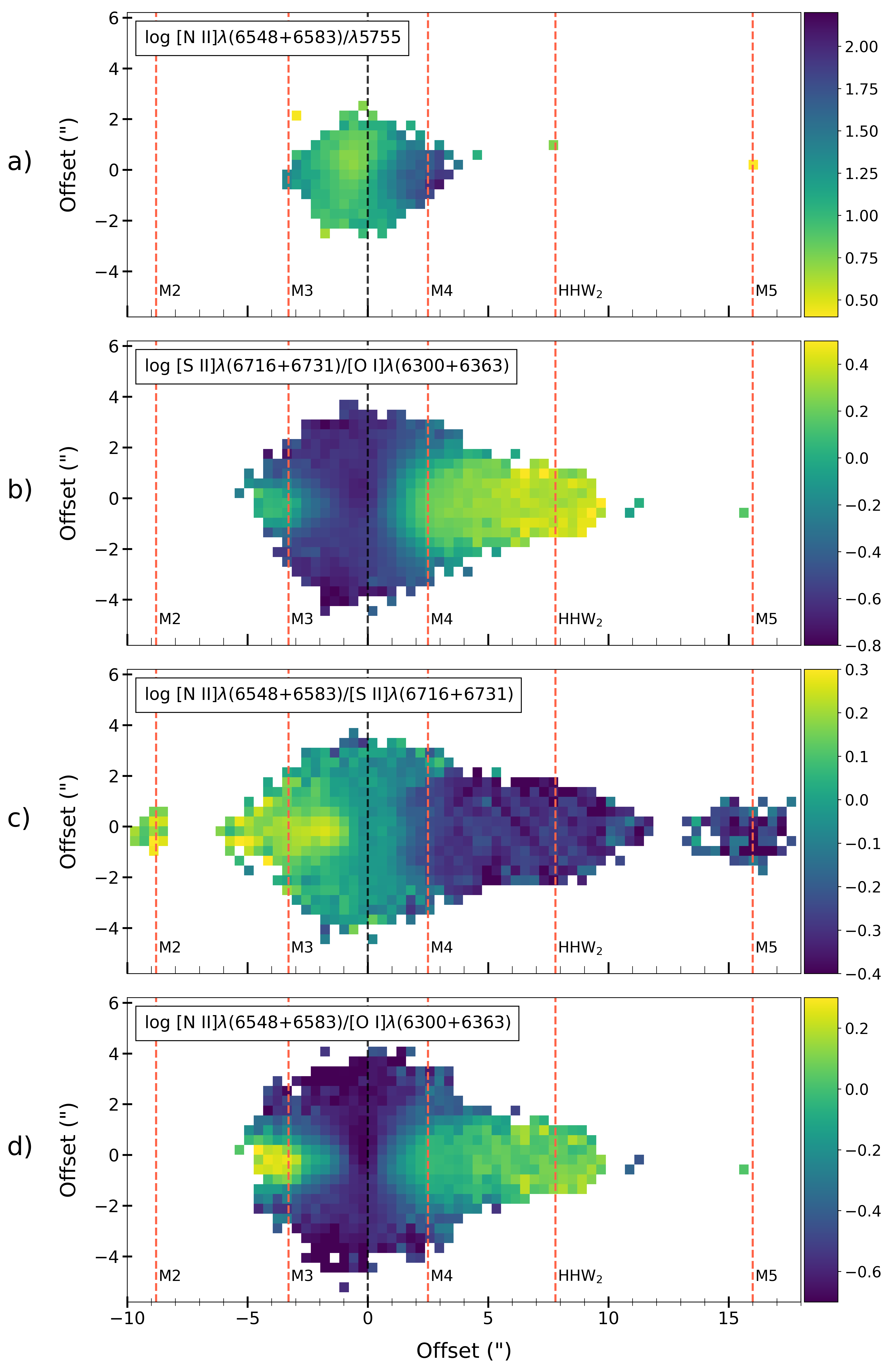}
        \caption{Maps from the MUSE data as in Fig. \ref{fig:MUSE_density}, for line ratios tracing \Te\ and \xe\ to show regions of higher excitation. Panel a), [\nii]$\lambda$(6548+6583)/5755 tracing electron temperature \Te; b), [\sii]$\lambda$(6716+6731)/[\oi]$\lambda$(6300+6363) tracing \Te\ and ionisation fraction \xe; c), [\nii]$\lambda$(6548+6583)/[\sii]$\lambda$(6716+6731) tracing \xe, \eden\ and \Te\; and d), [\nii]$\lambda$(6548+6583)/[\oi]$\lambda$(6300+6363) tracing \xe\ and \Te.}
        \label{fig:MUSE_excit}     
        \end{figure}

\begin{figure}
    \centering

    \includegraphics[width=9.5cm, trim= 0.5cm 1.0cm 2.3cm 0cm, clip=true]{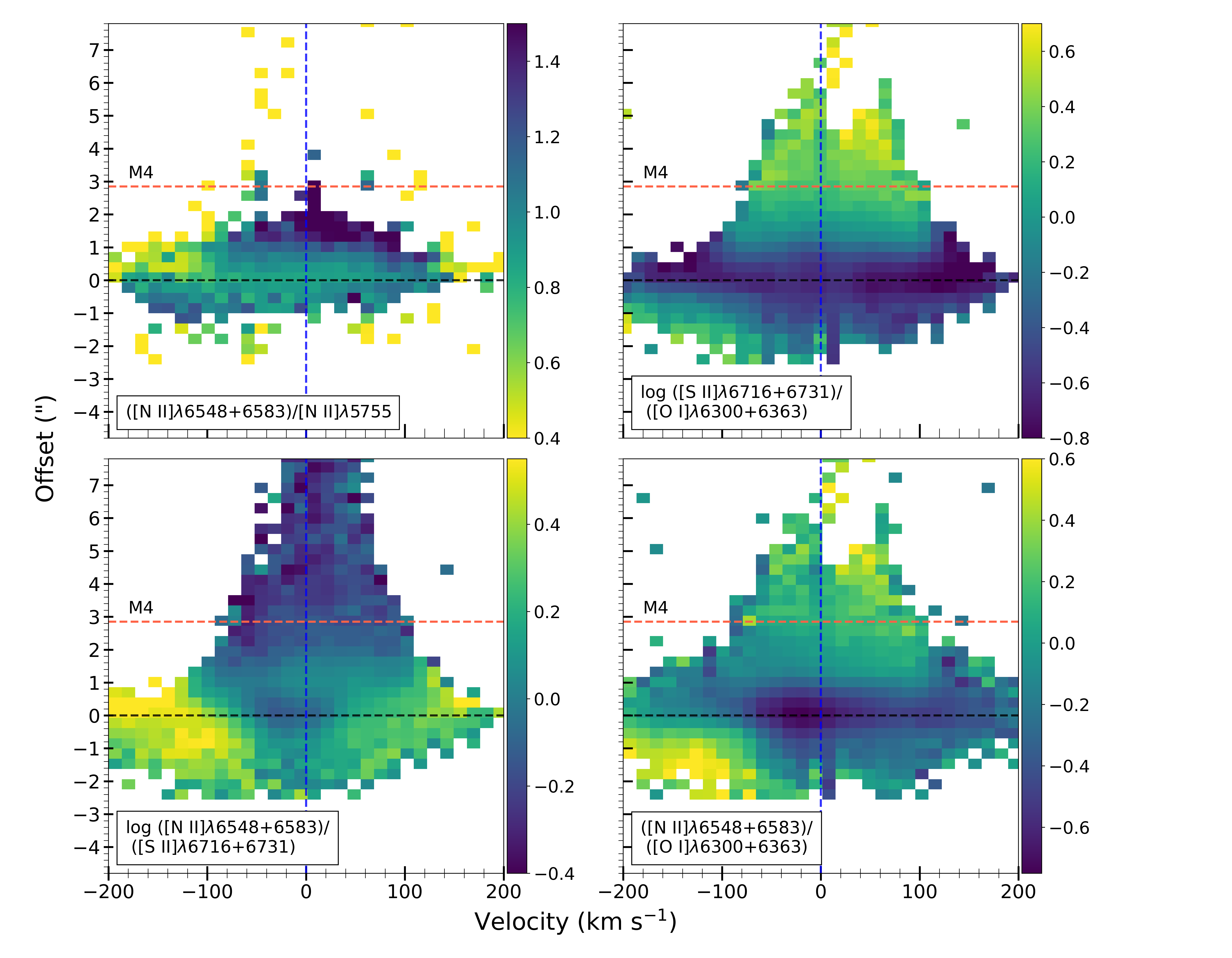}
        \caption{Position-velocity maps obtained from the X-shooter data as in Fig. \ref{fig:XS_ne}, for line ratios tracing \Te\ and \xe. The PV maps for each line ratio are dereddened along the jet axis and binned along both axes by a factor of 2. We note the high-excitation region observed at $<$ -100 \kms\ in the blue-shifted lobe (lower left within each panel) which likely represents the base of the blue-shifted jet. Panel a), [\nii]$\lambda$(6548+6583)/5755 tracing electron temperature \Te; b), [\sii]$\lambda$(6716+6731)/[\oi]$\lambda$(6300+6363) tracing \Te\ and ionisation fraction \xe; c), [\nii]$\lambda$(6548+6583)/[\sii]$\lambda$(6716+6731) tracing \xe, \eden\ and \Te\; and d), [\nii]$\lambda$(6548+6583)/[\oi]$\lambda$(6300+6363) tracing \xe\ and \Te.}
        \label{fig:XS_excit}     
        \end{figure}

  The ratio maps tracing primarily \Te\ and \xe\ are shown in Fig. \ref{fig:MUSE_excit}. These consistently indicate higher ionisation and temperatures in the blue-shifted lobe. Both lobes also appear to show a jump in ionisation fraction and temperature at approximately the position of the closest knot, suggesting shock excitation. The corresponding PV maps are shown in Fig. \ref{fig:XS_excit}. The [\nii]/[\sii] and [\nii]/[\oi] ratios in particular suggest higher excitation in the blue-shifted lobe. The region of high-velocity jet emission (bottom-left panels in the figures) is only visible within 2\arcsec, but shows markedly higher \xe\ and \Te\ than the red-shifted lobe. This asymmetry is not as clear in the [\sii]/[\oi] and [\nii] ratios, but these do show an increase in excitation in the red-shifted lobe from $\sim$1\arcsec\ ahead of the knot M4. The higher velocity resolution of these maps also allows us to detect a velocity trend in the red-shifted jet, with each of these ratios increasing from lower to higher red-shifted velocities indicating higher temperature and excitation.

    \subsection{Diagnostics using the BE method}
    
        \subsubsection{Method overview}    
        \label{subsection:bemethod}
     The BE method employs the prominent optical jet-tracing lines [\oi]$\lambda$$\lambda$6300, 6363, [\nii]$\lambda$$\lambda$6548, 6583, and [\sii]$\lambda$$\lambda$6717, 6731. Its application is relevant to regions without nearby strong sources of photoionisation, where S and N are singly ionised. The technique does not itself account for reddening of the line fluxes, which would affect the relevant ratios between them. However, an advantage of the BE method is that it makes use of lines close in wavelength (across a range of $<$ 450 \AA) and therefore the effect of reddening is minimised. It is therefore a useful technique to estimate physical diagnostics in jets where the extinction is not well-constrained. \citet{BE99} estimated the impact of reddening on the obtained values and conclude that at a value of A$_{v}$ = 3 mag, the introduced uncertainty on \xe\ and \Te\ is only 10-15$\%$, and so the impact is likely to be small especially in jets with low extinction. Similarly, an uncertainty of approximately 15$\%$ is associated with the commonly adopted uncertainties for the assumed local abundances relative to H \citep{Podio2006b}.
     
    The main uncertainties associated with this method then lie in the estimated line fluxes as well as its dependence on the adopted abundances and atomic parameters. A key limitation is that the density can only be reliably estimated below the critical density of the [\sii] lines ($\sim$ 2.5 $\times$ 10$^{4}$ \percm). However, this and the other assumed conditions are generally applicable within stellar jets that are known to emit FELs. Thus, this technique is a valuable tool to estimate physical diagnostics of jets in the optical band and well suited to analyses of large datasets.

    In the present analysis, we  used  an updated numerical code to apply the BE method, comparing the input line ratios with those predicted from a grid of possible values of \eden, \xe\, and \Te\ (see further discussion in \citet{Podio2006b,Melnikov2008}) that calculates the surfaces for \siirat, [\oi]/[\nii] and [\sii]/[\oi] at the same time. It then determines the values of \eden, \xe, and \Te\ that simultaneously match the model with the observations.  
    
    This approach has the advantage of including the effect of \Te\ variations in the estimate of \eden. An accurate estimate of \eden\ in turn has the effect of improving the diagnostic determination of all the parameters in regions of high ionisation and excitation. The code also makes use of abundances and collision strengths which are updated from the previous BE analysis of Th 28 \citep{BE99}. In particular, we assumed solar photospheric abundances of S, N, and O from \citet{Asplund2005}, which are lower than the values taken in the previous analysis.
    
    The blue-shifted lobe is less suited to analysis with this method due to both its faintness and its high ionisation, as indicated by the detection of [\oiii] emission in this lobe \citep{Murphy2021}. Although [\oi] emission can be detected to a distance of several arcseconds along the axis, as shown in Sect. \ref{subsection:oi_emission}, only a small fraction of this emission can be definitely attributed to the jet itself. We therefore concentrated the bulk of our diagnostic analysis on the red-shifted jet lobe, but we attempted to apply the method to the HV blue-shifted emission using the MUSE data. 

    \subsubsection{Results from the BE method}
        \label{subsection:be_results}

    \begin{figure*}
    \centering
        \includegraphics[width=18cm, trim=0cm 0cm 0cm 0cm, clip=true]{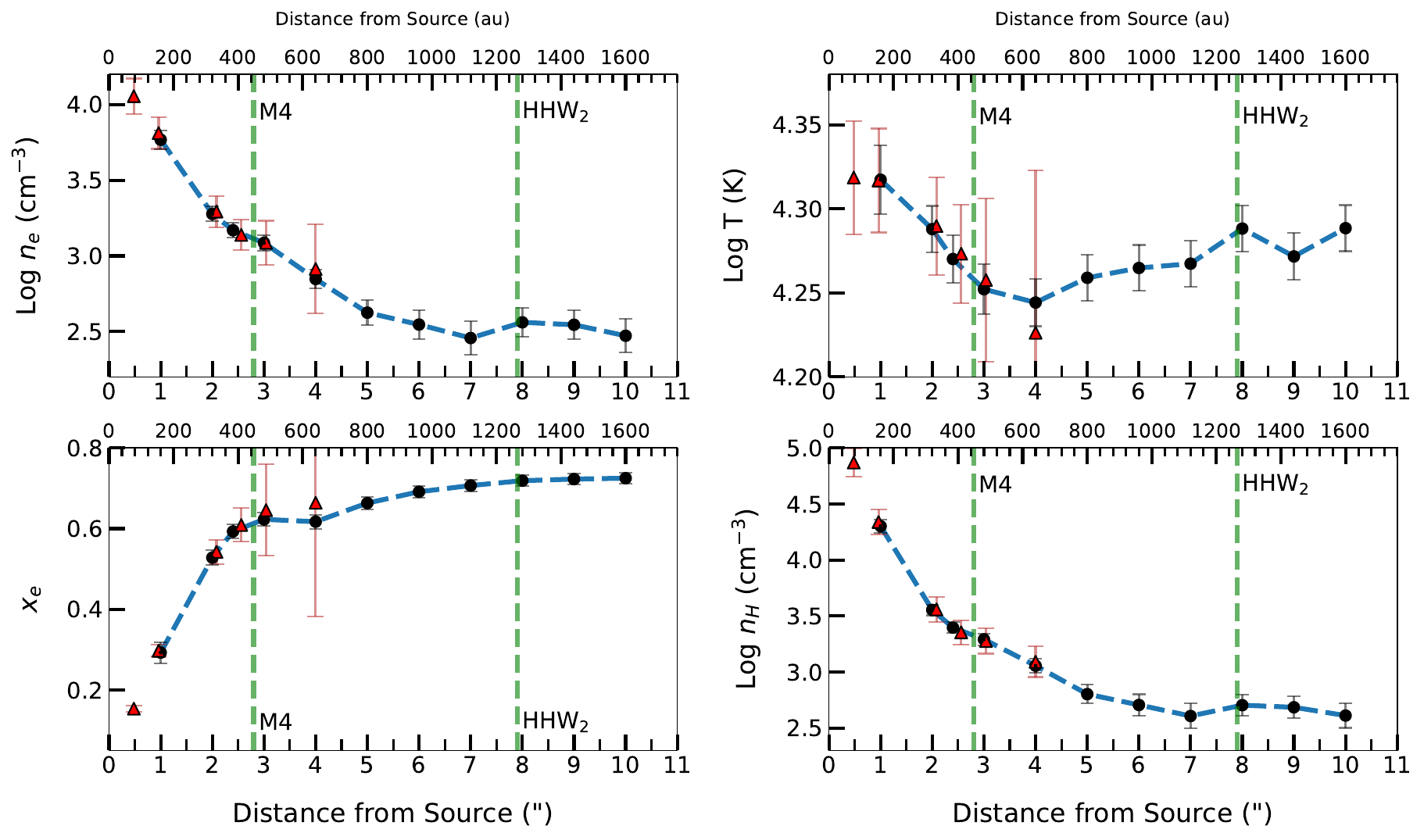}
        \caption{Results of the BE analysis from spectra sampled along the red-shifted jet axis. Clockwise from top left, the plots show electron density, \eden, electron temperature, \Te, hydrogen density, \nh\ ($= n_{e}/x_{e}$), and ionisation fraction, \xe. Results from the MUSE data are shown as black circles with the results from the X-shooter data shown as red triangles, and approximate knot positions are shown as green dashed lines. The \eden, \Te, and \nh\ values are plotted on a log scale. Error bars are derived from the flux calibration uncertainty as described in Sect. \ref{subsection:be_results}.}
        \label{fig:BE_profile}     
        \end{figure*}
    
    \begin{figure}
    \centering
        
     \includegraphics[width=9cm, trim= 0cm 0.2cm 2cm 0cm, clip=true]{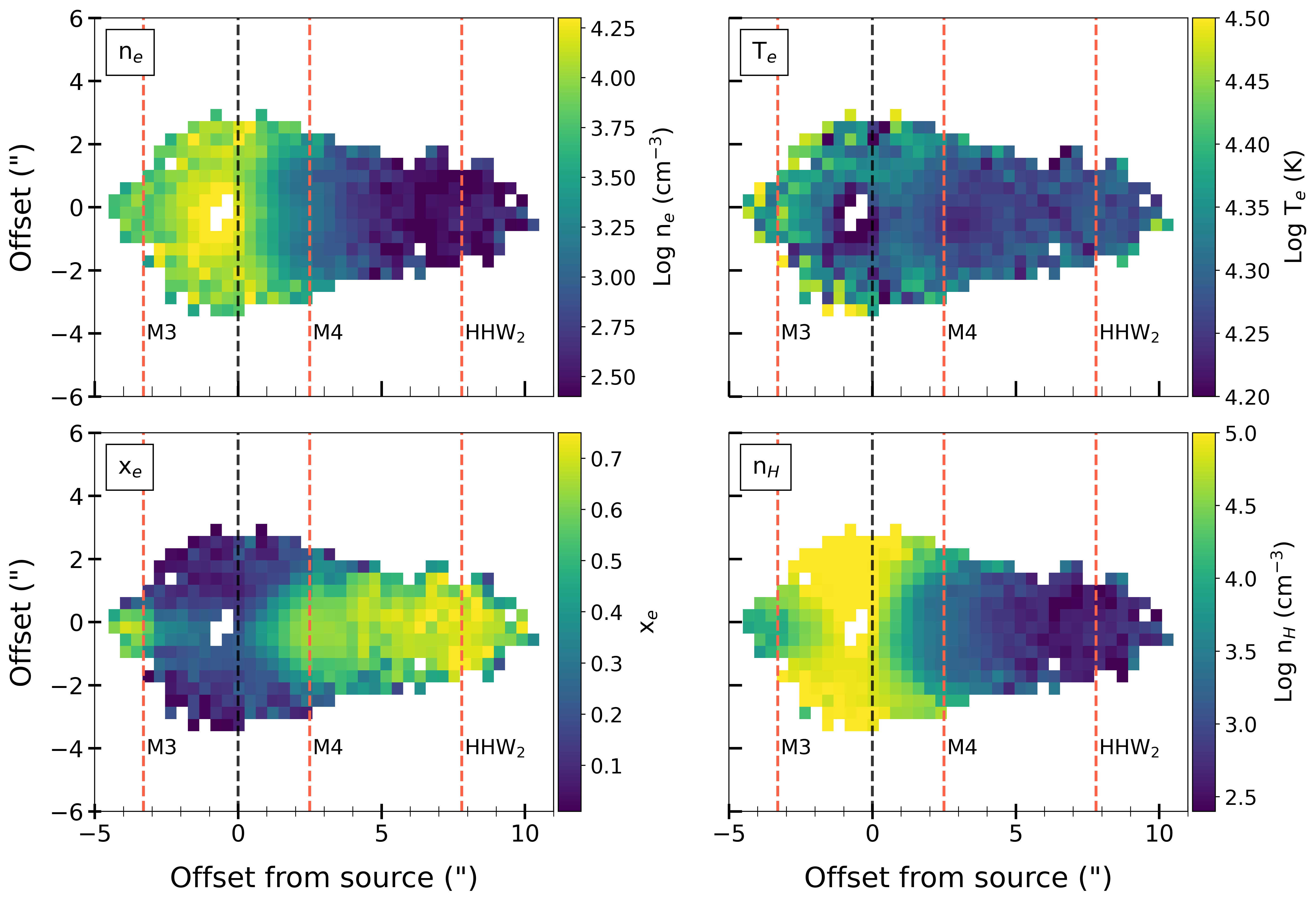}
        \caption{Spatial maps of the BE diagnostic results, obtained from the MUSE channel maps binned across the full red-shifted jet velocity range (-30 \kms\ to +140 \kms). The source position is shown by the black dashed line; red dashed lines mark the knot positions.}
        \label{fig:BE_redjet_channelmaps}     
        \end{figure}
 
    \begin{figure}
    \centering

        \includegraphics[width=9cm, trim= 0cm 0.2cm 1cm 0cm, clip=true]{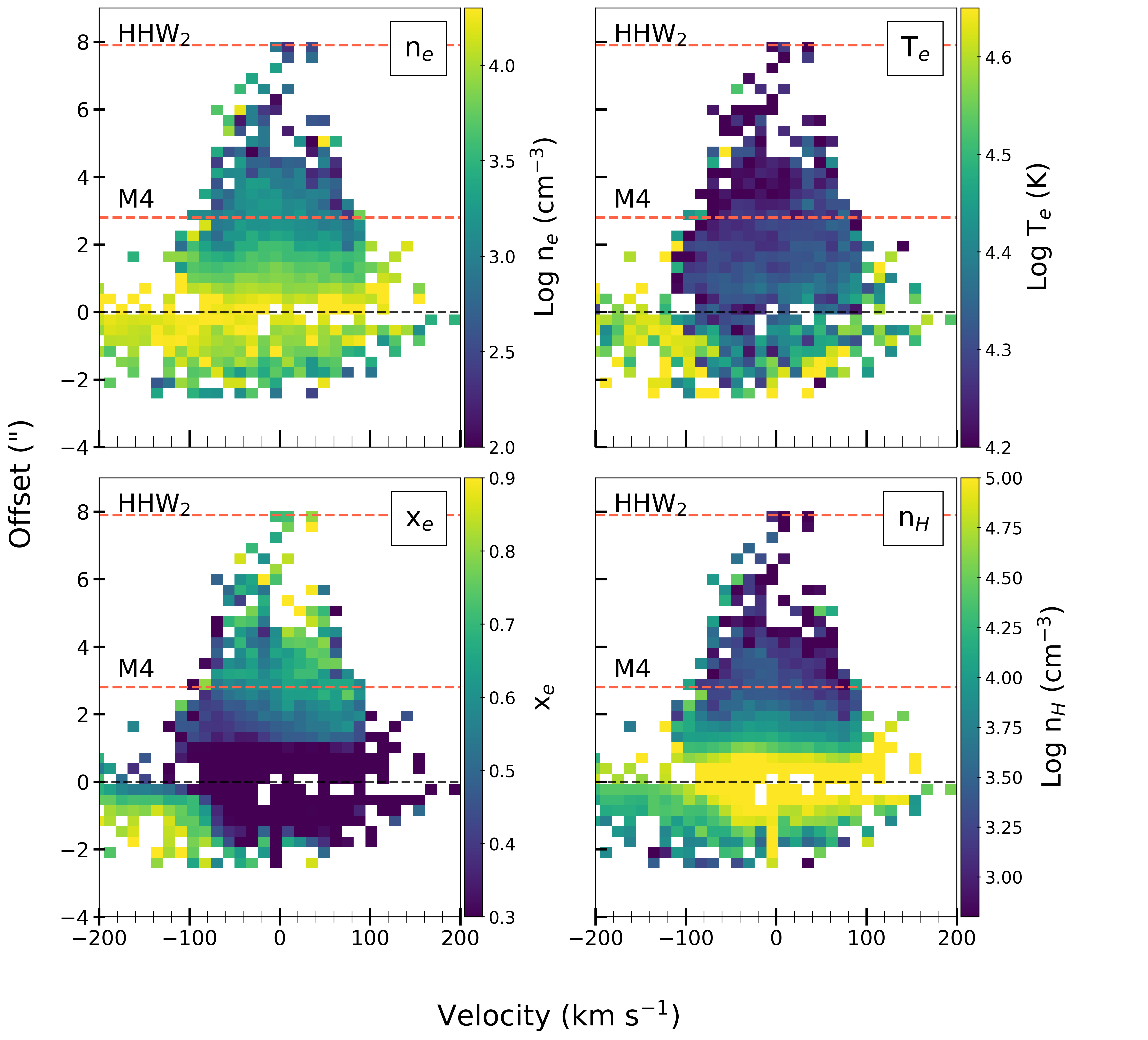}
        \caption{As in Fig. \ref{fig:BE_redjet_channelmaps}, for PV maps obtained from the X-shooter data.}
        \label{fig:BE_PVmaps}     
        \end{figure}

    We obtained line fluxes from 1D spectra of the relevant FELs, binned in 1\arcsec\ $\times$ 1\arcsec\ boxes at several positions along the jet axis. We carried out this procedure for both the MUSE and X-shooter data. Figure \ref{fig:BE_profile} shows the resulting profiles of the density, temperature, and ionisation fraction along the red-shifted jet; the results for knots M4 and \hhw\ are summarised in Table \ref{table:BE_results_summary}. We obtained almost identical results from the two datasets, except at the last position sampled in the X-shooter data (at +4\arcsec), where the [\oi] flux begins to drop sharply. To estimate the approximate uncertainty in these results associated with the flux calibration, we varied the input fluxes to the numerical code by the corresponding amount (5$\%$ for MUSE and 10$\%$ for the X-shooter measurements) for each of the positions shown in Fig. \ref{fig:BE_profile}. We then examine the spread of the resulting outputs. For the MUSE data, we therefore estimated typical uncertainties of 25$\%$ on \eden\ and \nh, 10$\%$ on \Te\, and 20$\%$ on \xe. Excluding the last position in the X-shooter data, where the \Ox\ flux becomes faint, we found corresponding estimates of approximately 35$\%$ for \eden\ and \nh, with 20$\%$ for \Te\ and 40$\%$ on estimates of \xe, reflecting the larger flux calibration error.
        
    We iterated this analysis over the pixels of the MUSE channel maps to obtain spatial maps of these parameters, excluding pixels  falling below a minimum flux threshold of 1 $\times~10^{-18}$ erg s$^{-1}$ cm$^{-2}$. This is shown in Fig. \ref{fig:BE_redjet_channelmaps}, with the channel maps binned over velocities from -30 \kms\ to +140 \kms\ to cover the line width of the red-shifted jet emission. We also obtained maps for the velocity channels between -200 to -30 \kms, covering the range where the [\oi] emission is most likely to be due to the blue-shifted jet itself (see Sect. \ref{subsection:oi_emission}). Only a small region of emission in the inner jet is above the flux threshold at these velocities. We sampled the mean values of each parameter from  1\farcs2 $\times$ 1\farcs2 boxes centred at -2\arcsec\ from the source. Because of the low inclination angle of the jet, the velocity resolution of the MUSE spectro-images is less ideal for exploring the jet properties in different velocity channels in comparison with the X-shooter spectra. Therefore, as shown in Fig. \ref{fig:BE_PVmaps}, we  obtained PV maps from the X-shooter data and iterated over them similar to the spatial maps.       
    
     Despite the caveats associated with applying the BE method to the blue-shifted jet lobe, these maps also support the asymmetries seen in the line ratio maps. Along the red-shifted jet, we found very high values of \eden\ ($>1 \times~10^{4}$ \percm) close to the source position, decreasing sharply along the outflow axis with small increases at the knots M4 and \hhw.  The hydrogen density, \nh\, follows a similar trend, reaching values of $>~5~\times~10^{4}$ \percm\ close to the source and decreasing sharply with small bumps at the knot positions. We also note the high-density hour-glass shaped feature around the source position and most prominent in the \nh\ map, which may be due to reflected emission from a cavity around the jet base. From the blue-shifted channel maps at -2\arcsec\ from the source, we found \eden\ = 4.5 $\times~10^{3}$ \percm\ and \nh\ = 6.9 $\times~10^{3}$ \percm, roughly twice the values at the corresponding position in the red-shifted jet.

    As shown from the results of the BE diagnostics in Fig. \ref{fig:BE_profile}, the electron temperature in the red-shifted lobe decreases slightly moving outwards from the source, from 2 $\times~10^{4}$ K to 1.8 $\times~10^{4}$ K just after the position of the knot M4; it then increases slowly with a small peak at \hhw. The ionisation fraction, \xe\,, is very low ($< 0.1$) close to the source and rises sharply until it reaches a value of 0.6 at the position of M4, and then increases slowly to 0.7 at \hhw. This is also illustrated in the channel maps in Fig. \ref{fig:BE_redjet_channelmaps}, which also show the highest ionisation centred along the jet axis. At -2\arcsec\ in the blue-shifted jet we obtain \Te\ = 4.2 $\times~10^{4}$ K, \xe\ = 0.67 as compared with \Te\ = 1.9 $\times~10^{4}$ K and \xe\ = 0.53 at the same position in the red-shifted jet. This indicates that while the blue-shifted lobe indeed is indeed hotter and more ionised, \nh\ is also still higher than in the red-shifted lobe.

        \begin{table*}
        \centering
        
        \caption[target]{Diagnostics at the knot positions in the red-shifted jet lobe.}
        \renewcommand{\arraystretch}{1.8}
        \begin{tabular}{{p{0.06\textwidth}<{\raggedright} p{0.07\textwidth}<{\raggedright} p{0.06\textwidth}<{\raggedright} p{0.05\textwidth}<{\raggedright} p{0.12\textwidth}<{\raggedright} p{0.07\textwidth}<{\raggedright} p{0.05\textwidth}<{\raggedright} p{0.05\textwidth}<{\raggedright} p{0.09\textwidth}<{\raggedright}}}
                \hline \hline
             & \multicolumn{4}{l}{MUSE} & \multicolumn{4}{l}{X-shooter} \\
             \hline
                Knot &  \makecell[l]{\eden\ \\ (\percm)} &  \makecell[l]{\Te\ \\ (10$^{4}$ K)} & \xe\ & \makecell[l]{\nh\ ($= n_{\mathrm{e}}/x_{\mathrm{e}}$) \\(\percm)} & \makecell[l]{\eden\ \\ (\percm)} & \makecell[l]{\Te\ \\ (K)} & \xe\ & \makecell[l]{\nh\ ($= n_{e}/x_{e}$) \\(\percm)} \\
                M4    & 1480 & 1.86 & 0.59 & 2508 & 1377 & 1.88 & 0.6 & 2295 \\
                \hhw\ & 364  & 1.94 & 0.72 & 506  &  & & & \\ 
                \hline
        \end{tabular} 
        \label{table:BE_results_summary}
        \end{table*}
    
    \subsection{Transverse maps}
    \label{subsection:tv_maps}
    
\begin{figure*}
    \centering
        \includegraphics[width=17cm, trim= 0cm 0cm 0cm 0cm, clip=true]{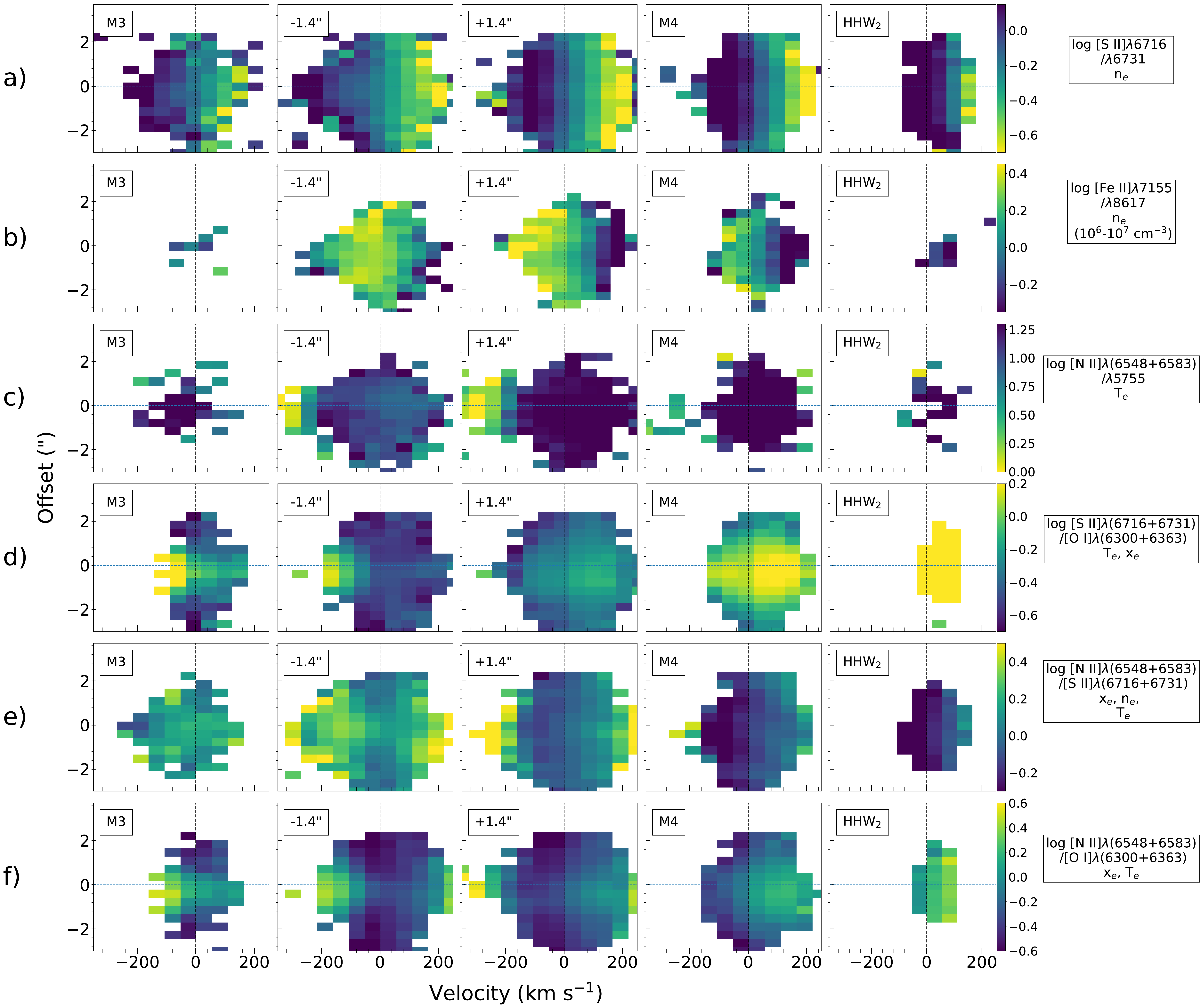}
        
        \caption{Transverse PV maps across the Th 28 jets, sampled from the MUSE data at knot positions and at +/-1.4\arcsec\ from the source position (close to the jet base). The maps are binned across 1\arcsec\ sections along the jet axis and colour bars are as in Fig. \ref{fig:MUSE_density}. Rows a) and b) show the \siirat\ and \ferat\ ratios, both primarily sensitive to electron density \eden. The remaining rows are as follows: c), [\nii]$\lambda$(6548+6583)/5755 tracing electron temperature \Te; d), [\sii]$\lambda$(6716+6731)/[\oi]$\lambda$(6300+6363) tracing \Te\ and the ionisation fraction \xe; e), [\nii]$\lambda$(6548+6583)/[\sii]$\lambda$(6716+6731) tracing \xe, \eden\ and \Te\; and f), [\nii]$\lambda$(6548+6583)/[\oi]$\lambda$(6300+6363) tracing \xe\ and \Te. }
        \label{fig:combined_tv_rmaps}     
        \end{figure*}

\begin{figure*}
    \centering
        \includegraphics[width=14cm, trim= 0cm 0cm 0cm 0cm, clip=true]{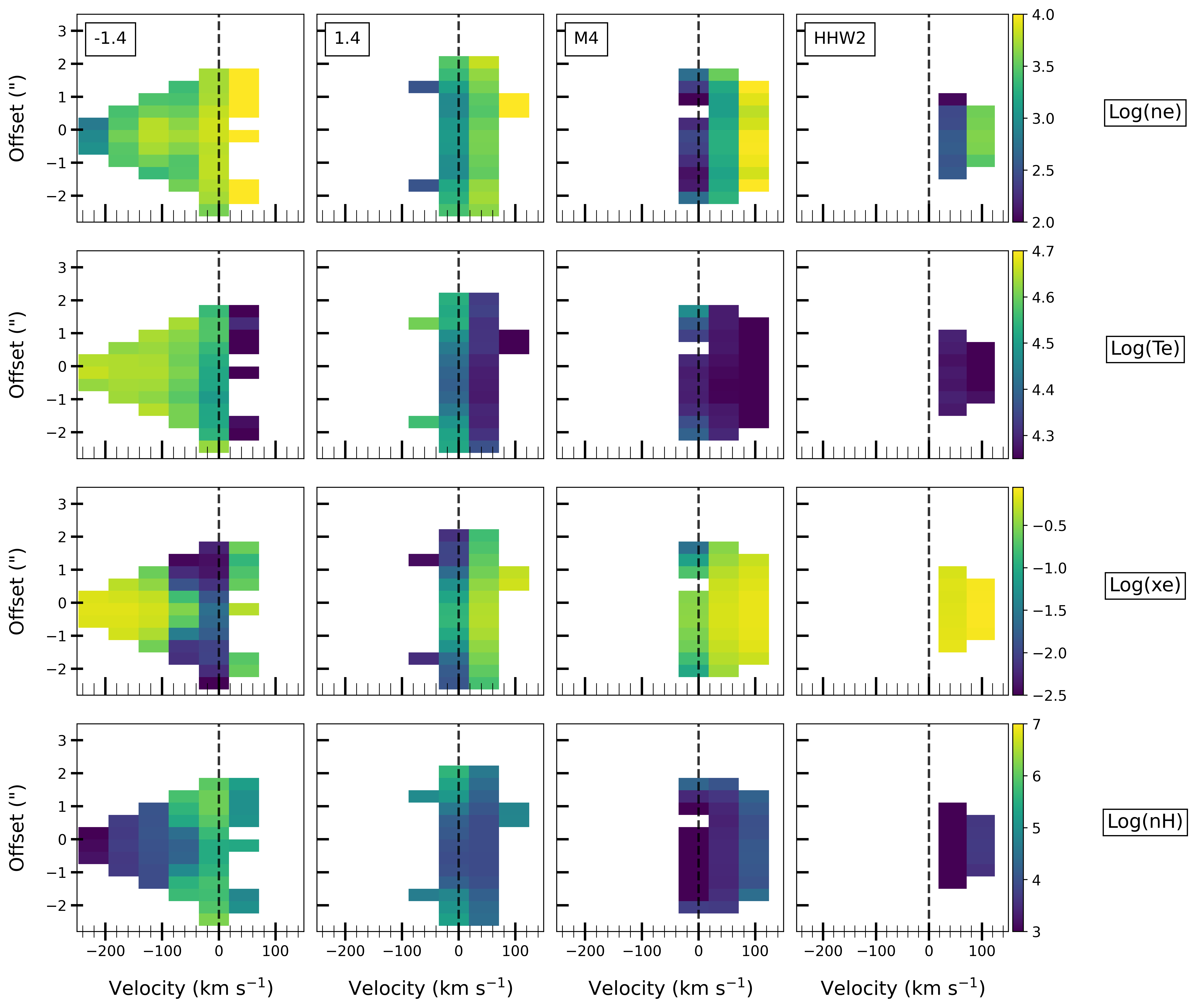}
        \caption{Transverse PV maps from the MUSE data, showing the BE analysis results for the red-shifted jet knots. colour bars are as in Fig. \ref{fig:BE_redjet_channelmaps}.}
        \label{fig:BE_TV_red}     
        \end{figure*}

        The MUSE datacubes allow us to further explore the jet line ratios by taking PV maps orientated across the jet axis. This allows us to look for transverse asymmetries in the position-velocity space similar to those observed by \citet{Coffey2008}. These  maps were sampled from 1\arcsec\ sections of the jet axis, centred at knot positions and also two positions +/-1\farcs4 from the source (i.e., near the base of the jet emission in both lobes). The resulting line ratio maps are shown in Fig. \ref{fig:combined_tv_rmaps}, with the three left-hand columns showing cuts across positions in the blue-shifted jet and the four right-hand columns showing cuts from the red-shifted jet. As in the previous section, we further apply the BE method to derive transverse maps of the jet parameters including \nh. Figure \ref{fig:BE_TV_red} shows the BE results obtained at the red-shifted knot positions as well as at -1\farcs4 and +1\farcs4 from the source (the inner regions of the blue- and red-shifted lobes, respectively).

    The two upper rows of Fig. \ref{fig:combined_tv_rmaps} show the \siirat\ and [\feii] ratios. In the red-shifted jet, we find that the [\feii] ratio is only detected at the jet base and in the closest knot M4. However this shows a pronounced velocity trend with higher \eden\ in the lower velocity channels. In contrast, the \siirat\ ratio shows decreased values (i.e., higher \eden) at higher velocities in all of the red-shifted knots. Since the [\feii] ratio is sensitive to regions of higher \eden, this may suggest it is tracing a lower-velocity, higher-density component, while \SII\ primarily traces the faster jet core. We also note that the [\sii] ratio may be subject to the same caveats discussed for [\oi] in Sect. \ref{subsection:oi_emission}, with multiple low-velocity emission components seen close to the jet base. The BE maps generally follow a similar trend to \siirat,\ as this is one of the ratios used in this analysis; due to the low inclination of the red-shifted jet and the flux threshold required for the numerical code to converge, the different velocity components are not clearly differentiated. However, we note that in most of the knots, both \eden\ and \nh\ increase slightly in the higher velocity bins.

     At the blue-shifted jet positions, the knots M2 and M3 are only observed in [\sii], which shows higher \eden\ in lower-velocity blue-shifted channels. M3 is detected both in high-velocity blue-shifted channels and in red-shifted velocity channels up to $\sim 100$ \kms, and is likely a bow shock similar to the other knots in this lobe. The blue-shifted jet base shows the same trends as the red-shifted jet base in both of these line ratios, suggesting that close to the source position both ratios may be contaminated by scattered emission from the red-shifted jet.

    The lower four rows of Fig. \ref{fig:combined_tv_rmaps} show the line ratios tracing excitation.  In the red-shifted jet, we find that the knot M4 also shows slightly higher \Te\ in the lower velocity channels. The low velocity resolution of the MUSE data and the low jet inclination likely make the velocity components of the red-shifted jet poorly separated in these maps. Despite this, the knot positions also consistently show higher \xe\ values in the higher-velocity red-shifted channels. The knots also show a general increase in temperature and \xe\ with distance from the source.
    
    In the blue-shifted jet, M2 is faint and is observed only in one excitation-tracing ratio, [\nii]/[\sii]. The inner knot M3 is observed in several line ratios and shows a slight increase in excitation at high velocities, but it primarily shows higher \Te\ and \xe\ in the spatial center. In both the line ratio maps and in the BE method maps (Fig. \ref{fig:BE_TV_red}, left column) we see a better differentiation of the velocity components than in the red-shifted jet. Lower-velocity channels appear to show higher \eden\ and \nh, but lower \Te\ and \xe, though the channels at low velocities $\simeq$ 0 \kms\ are likely to contain significant contributions from scattered light and the LV [\sii] and [\oi] component (discussed in Sect. \ref{subsection:oi_emission}). Nevertheless, this trend appears continuous with the higher-velocity channels $>$ 50 \kms; these are more reliably associated with the jet and show lower densities with higher temperatures and ionisation fraction. 

    The transverse cuts across the inner parts of the jets show consistent axial symmetry: the spatial centre of the jet in both lobes shows higher \xe\ and \Te, decreasing from the axis to the jet edges. We note that the spatial resolution in these maps is 0\farcs9  compared with the observed extent of the line emission +/-2\arcsec\ from the jet axis, which may result in a smoothing of the observed trends in this direction. In the ratio maps this is especially visible in [\sii]/[\oi]. Along with the lower \nh\ values observed in the same regions, this is consistent with higher temperatures and ionisation in the spatial centre and the higher-velocity channels; this follows the expectations if these do indeed trace the faster, hotter jet core. While these findings are hampered by poor S/N in the blue-shifted lobe and poor velocity differentiation in the red, both the spatial symmetry and velocity trends in the jets tend to suggest a hotter, less dense, and more ionised core surrounded by cooler and denser outflow layers.  

    \subsection{Accretion rates}
        \label{subsection:macc_estimates}
   
    We estimated the accretion luminosity, \Lacc\, and, hence, the mass accretion rate, \Macc\, from the spectrum at the source position, using emission lines whose luminosity, $L_{\mathrm{line}}$, is known to be well correlated with the mass accretion. However, the edge-on disc of Th 28 presents a challenge in this regard, as a clear stellar spectrum is not observed. Instead, the observed line fluxes are significantly reduced due to obscuration by the edge-on disc. Two main components should be considered: wavelength-dependent extinction and 'grey scattering.' The former is discussed in more detail in Appendix \ref{section:extinction}. Grey scattering is wavelength independent and can be approximated as an obscuration factor that suppresses \Lacc\ and \Lstar\ by the same amount. As such, it has no effect on the relationship of \Lacc\ to $L_{\mathrm{line}}$, and once it is estimated, it can be corrected for in a straightforward way. From \citet{Alcala2014}:

    \begin{equation}
        \dot M_{\mathrm{acc,corrected}}~ = (\mathrm{obscuration~factor})^{1.5}\dot M_{\mathrm{acc, observed}}
    .\end{equation}
    
    The obscuration factor for Th 28 was not previously estimated, but is expected to be substantial due to the edge-on disc and the under-luminosity of the source. However, we can make an estimate of the obscuration using the \Ox\ line. This line is emitted from the jet, so that the emission originates above the dust plane and is theoretically less subject to obscuration, but is also indirectly correlated with the accretion rate \citep{Herczeg2008,Riaz2015}. We can therefore use this tracer as a point of comparison to estimate the approximate mean obscuration as discussed below.
    
    The extended wavelength range of the X-shooter spectra allows us to access a broader array of accretion tracers than can be measured from the MUSE data alone, while directly comparing \Macc\ estimates from the tracers common to both datasets. For each one a continuum-subtracted spectrum was extracted in a 1\arcsec\ x 1\arcsec\ box centred on the source position in each cube. The estimated $L_{\mathrm{line}}$ from each spectrum was extinction-corrected for the averaged value of $A_{v}$ $\sim 2.5$ mag across the source region based on the function obtained in Appendix \ref{section:extinction}. We then derived \Lacc\ from the relationship:
    
    \begin{equation}
        \log L_{\mathrm{acc}} = \mathrm{A} \log L_{\mathrm{line}} + \mathrm{B} 
    ,\end{equation}
    
    with coefficients A and B determined empirically from correlation of line fluxes with accretion luminosity \Lacc\ \citep{Herczeg2008, Alcala2014}.  For the permitted lines, we employed the coefficients obtained by \citet{Alcala2017}; whereas for \Ox,\ we used the updated coefficients for the HVC from \citet{Nisini2018}. Due to the low inclination of this jet, we could not resolve a separate HVC and LVC, particularly in the MUSE spectra. Therefore, we used the relationship obtained for the HVC based on the assumption that the emission in this line is dominated by the bright high-velocity jet, which is less susceptible to obscuration.  The accretion rate is then given by: 
    
    \begin{equation}
        \dot M_{acc} = \dfrac{1.25 L_{\mathrm{acc}} R_{*}}{G M_{*}}
    .\end{equation}
    
    We estimated \Rstar\ from the stellar luminosity as follows:
    
    \begin{equation}
        L_{*} = 4 \pi R^{2}_{*} \sigma T^{4}_{\mathrm{eff}}
    .\end{equation}
    
   From \citet{Alcala2017}, we obtained the stellar parameters of  $T_{\mathrm{eff}}$ = 4900 K and  \Lstar\ = 2.0 \Lsun, taking \Mstar\ = 1.6 \Msun\ \citep{Louvet2016}. Based on these values, we have \Rstar\ $\sim$ 1.96 \Rsun. We thus obtained initial estimates of \Macc\ for each of our accretion tracers. We then obtained the average ratio between the value of \Macc\ derived from the \Ox\ flux and each of the permitted lines; then, we could derive an estimate of the obscuration factor. We use this obscuration factor to correct the values of \Macc\ from the permitted lines.
    
   The full list of accretion tracers in the MUSE and X-shooter spectra, along with the relevant coefficients, are provided in the supplementary material. Figure \ref{fig:Macc_combined} shows the log \Macc\ for the different lines with the permitted lines shown before and after correction for obscuration. In the MUSE data, we obtained an estimate of \Macc\ = 2.2 $\times~10^{-7}$ \Msun~\peryr\ from the [\oi] line luminosity and an average obscuration of 58. Combining this with the corrected values of \Macc\ using our obscuration factor, we obtained an average value of \Macc\ = 2.5 $\times~10^{-7}$ \Msun~\peryr. From the X-shooter spectra we similarly obtain an obscuration factor of 50 and average \Macc\ = 2.3 $\times~10^{-7}$ \Msun~\peryr. Hence, we found an overall average of \Macc\ = 2.4 $\times~10^{-7}$ \Msun~\peryr. We used this value in comparisons with the mass outflow rate to  account for the variability in the \Macc\ estimates using the full range of emission lines (see \citet{Whelan2014b}). 

    However, an examination of the results plotted in Fig. \ref{fig:Macc_combined} (left panel) reveals a small downward trend of \Macc\ with increasing wavelength, suggesting an over-correction for extinction. As discussed in Appendix \ref{section:extinction}, this may result from the emission at the source being more optically thin than our initial assumption. Thus, we also chose to calculate \Macc\ using the on-source \av\ of 1.26 mag obtained from the [\feii] lines, with the results shown in the right panel of Fig. \ref{fig:Macc_combined}. The full tables are provided in the supplementary material. These results no longer show a trend with wavelength. With this assumption, we obtained \Macc\ = 5.65 $\times~10^{-8}$ \Msun~\peryr\ and = 5.4 $\times~10^{-8}$ \Msun~\peryr\ from the MUSE and X-shooter spectra, respectively, yielding an average \Macc\ of 5.3 $\times~10^{-8}$ \Msun~\peryr. The corresponding obscuration factors are then 45 and 43. We discuss these results further in Sect. \ref{subsection:accretion_discuss}.

    \begin{figure*}
    \centering
        \includegraphics[width=9.0cm, trim= 0cm 0cm 0cm 0cm, clip=true]{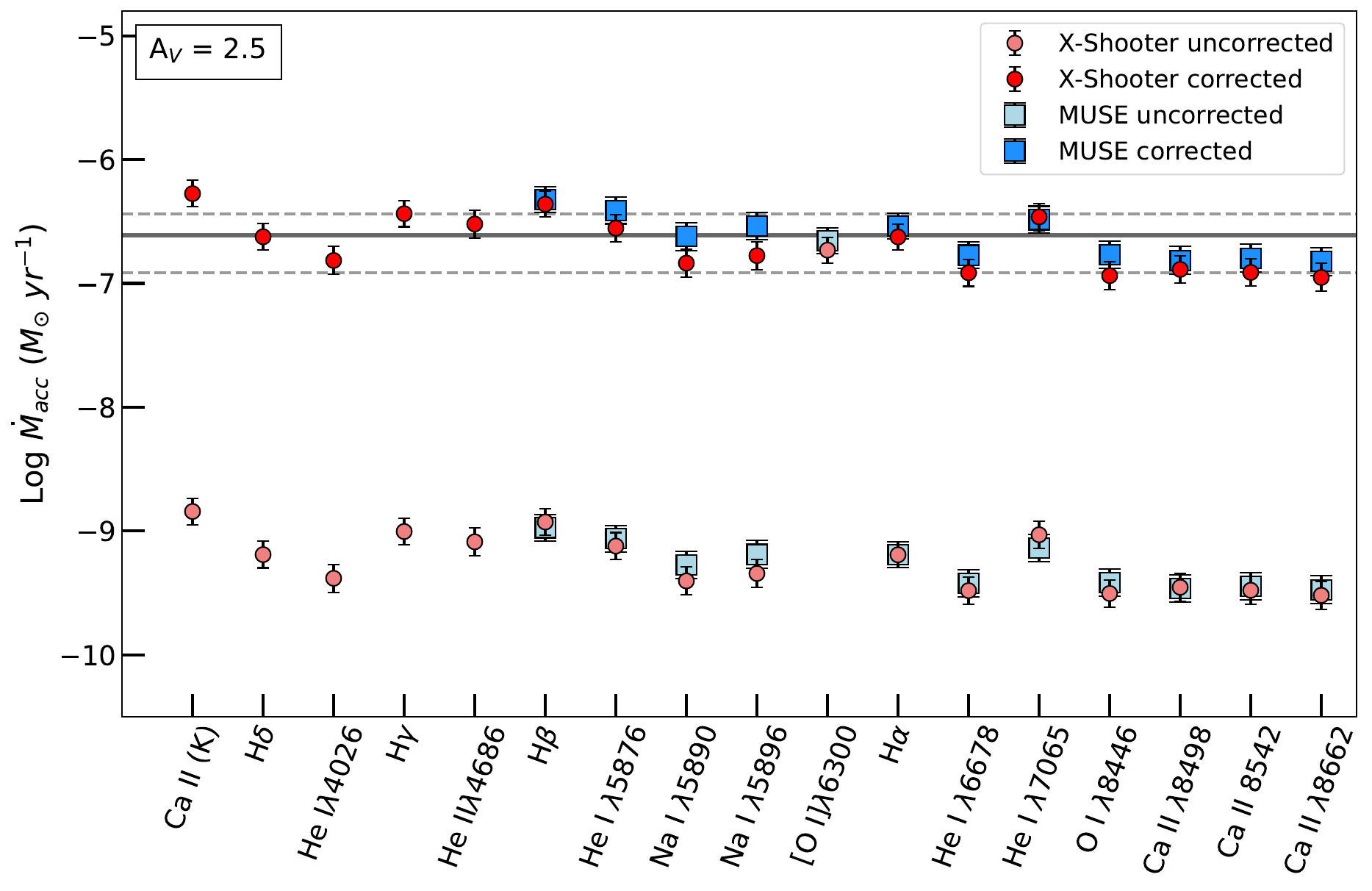} \includegraphics[width=9.0cm, trim= 0cm 0cm 0cm 0cm, clip=true]{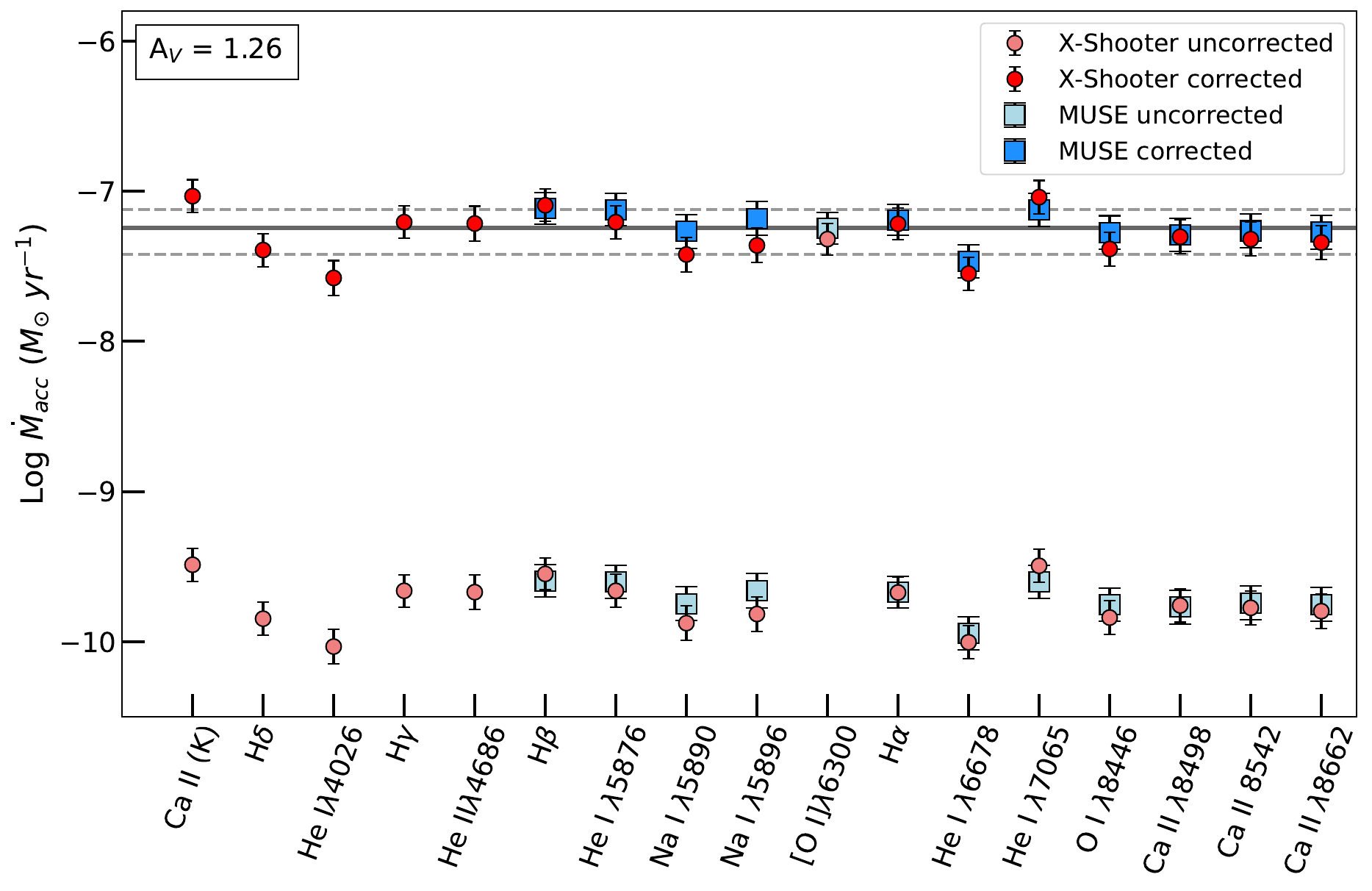}
        \caption{Combined estimates of \Macc\ from the MUSE (squares) and X-shooter (circles) spectra, showing values obtained for \av\ = 2.5 mag and 1.26 mag (left and right panels, respectively). The estimates are shown before and after correction for the estimated obscuration value in each case. The solid horizontal line shows the mean \Macc\ value, with the 1-$\sigma$ bounds shown by dashed lines. In the left panel a slight decrease is observed in \Macc\ with increasing wavelength, suggesting over-correction for extinction.}
        \label{fig:Macc_combined}     
        \end{figure*}  

    \subsection{Mass outflow and jet efficiency}
    \label{subsection:mass_outflow}

    We estimated the rate of mass outflow (\Mout) through the jet using two methods: 1) by combining the density of the jet estimated with the BE method and the estimated jet cross-section and 2) by using the line luminosities of \Ox\ and \siib,\ following the methods of \citet{Agra-Amboage2009}, assuming uniform gas conditions and volume emission over the aperture, and \citet{Bacciotti2011}, respectively. We estimated the jet FWHM from Gaussian fitting of the jet profile, binned over 1\arcsec\ sections along the jet axis to match the sampled spectra. We used the deconvolved [\sii] MUSE cubes to estimate the FWHM as described in \citet{Murphy2021}, and estimate the mass outflow rate as $\dot M_{out} = \mu m_{H} n_{H} \times \pi r^{2}v_{j}$. The width of the red-shifted jet ranges between 0\farcs9-1\farcs2 at the knot positions (144-192 au at 160 pc) and the deprojected jet velocity $v_{\mathrm{j,red}}$ is 270 \kms. The estimates of mass outflow using FEL luminosities also take into account electron density and ionisation fraction of the emitting gas, which were also obtained with the BE method at the sampled positions along the jet. We further compare the estimates based on the MUSE data with those obtained using the FEL fluxes estimated with X-shooter.

    For the values of \Mout\ obtained using \nh, we considered the uncertainties in the jet velocity, width, and the measured density. Due to the small inclination angle of the jet, the measurement of the velocity is dominated by the proper motions which are well-constrained in the red-shifted lobe; we therefore took an uncertainty of +/-0\farcs05 \peryr\ (or approximately 30 \kms\ \peryr) corresponding to an 11$\%$ uncertainty. We took a 10$\%$ uncertainty on the jet radius and a conservative 40$\%$ uncertainty on the jet density. This corresponds to an overall uncertainty of 40$\%$ on the estimated values of \Mout. We assume the cross-section volume is uniformly filled with gas at the estimated density (i.e., it has a filling factor = 1). This assumption may not hold as the jet travels farther from the source, but the regions of lower density included in the cross-section should be partly compensated for by regions of higher density than those traced by the FELs used in the BE method.
    
    Using the FEL luminosities to estimate \Mout, uncertainties arise due to the assumed distance, extinction, and the measured \eden\ and \xe\ derived from the BE method. In particular, the value of \xe\ has a strong impact on the value of \Mout\ derived from the \Ox\ line. The dominant consideration, however, is the uncertainty on the line fluxes, as this not only determines the estimated line luminosity but also affects the estimates of \eden\ and \xe\ derived from the optical line ratios. In both the MUSE and X-shooter data, these are dominated by the instrumental flux calibration uncertainties of 5$\%$ and 10$\%$, respectively. Taking into account the uncertainty in \eden, we obtain uncertainties of 50$\%$ for values of \Mout\ derived from the \Ox\ line luminosity and 40$\%$ using \siib.
    
\begin{figure*}
    \centering

    \includegraphics[width=16cm, trim= 0cm 0cm 0cm 0cm, clip=true]{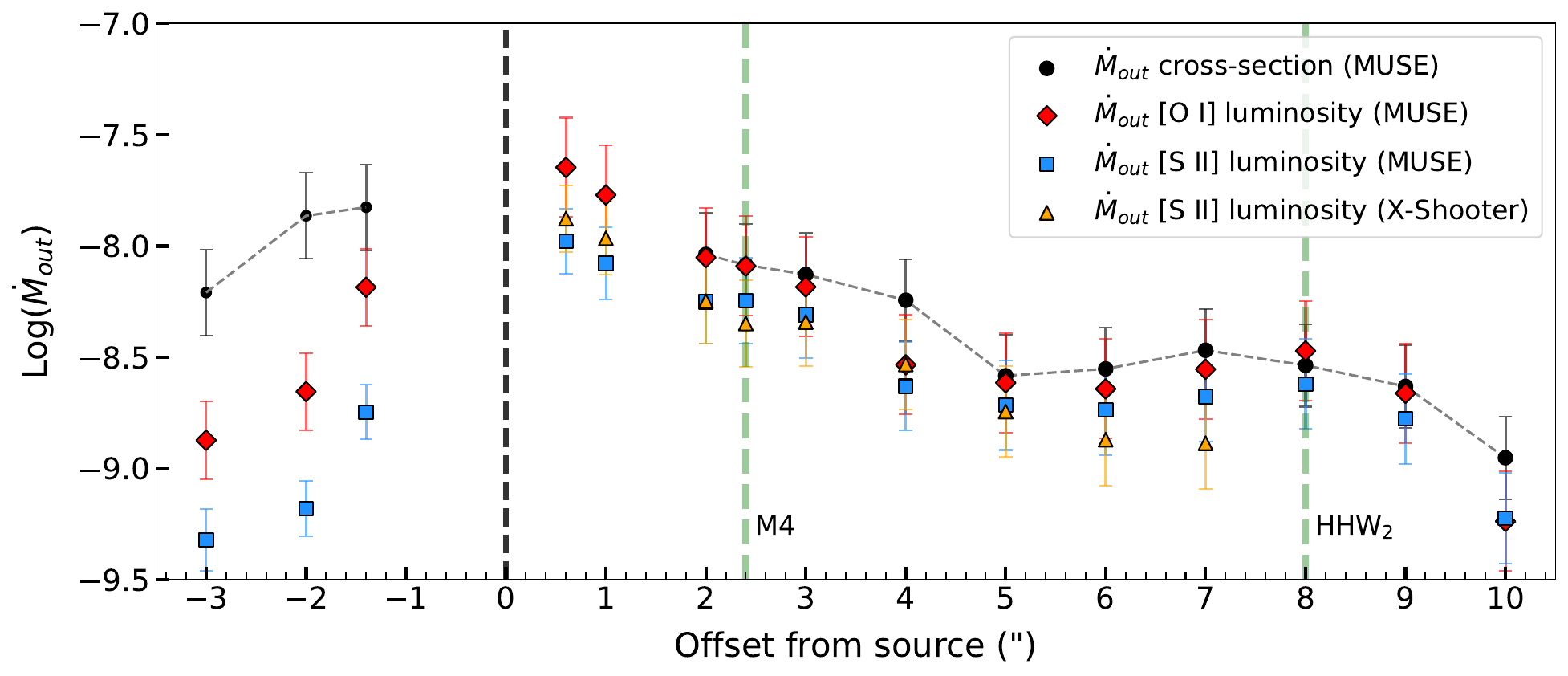}
        
        \caption{Mass outflow rates through the jets obtained with MUSE and X-shooter. Knot positions are marked with green dashed lines and the source position is marked by the black dashed line. The inner two spatial positions ($<$ 2\arcsec\ of the source) are excluded from the cross-sectional estimate in the red-shifted lobe since the jet width is unresolved in this region.} 
        \label{fig:th28_red_out}     
        \end{figure*}

    The full measurements for the red-shifted and blue-shifted jet lobes are shown in Fig. \ref{fig:th28_red_out}, while Table \ref{table:mass_outflow} shows the estimated mass outflow rates in both lobes. In the red-shifted jet lobe, estimates of \Mout\ range from 3 $\times~ 10^{-9}$ \Msun~\peryr\ to 5.8 $\times~10^{-9}$ \Msun~\peryr. Overall, the X-shooter line luminosities yield the lowest estimates of \Mout. 
    
    We also attempted to estimate \Mout\ in the fainter blue-shifted jet lobe, which shows [\sii] and weak \Ox\ emission (close to the source). Here, we have $v_{\mathrm{j,blue}}$ = 360 \kms\ from the proper motions. We used the values of \eden, \xe\ etc. obtained for the blue-shifted jet in Sect. \ref{subsection:be_results}, and obtain both jet widths and line fluxes from the same spectro-images used to isolate the blue-shifted emission in that section. Line fluxes in the red-shifted lobe are not extinction-corrected due to the low values of \av\ here; however, we corrected those in the blue-shifted lobe with the averaged \av\ function over this spatial region. We are able to estimate \Mout\ at -3\arcsec, -2\arcsec, and -1\farcs2 from the source. We note that the latter position is just within the region where the jet width is unresolved;  therefore, we took the width of the spatial PSF as an upper limit.
    
    In this lobe, we find values of \Mout\ ranging between 1.0 $\times 10^{-9}$ \Msun~\peryr\ to 1.2 $\times 10^{-8}$ \Msun~\peryr, which are reasonable values for a CTTS jet. Comparing similar distances along each of the lobes, \Mout\ measured with \nh\ is comparable or slightly lower in the red-shifted lobe with an average ratio of 0.72, due to the lower estimated \nh\ and the lower jet velocity (note that $v_{\mathrm{,red}}$ = 0.75$v_{\mathrm{t,blue}}$). In contrast, the estimates of \Mout\ via line luminosity are both substantially lower in the blue-shifted jet, by average factors of 4 and 10 using the [\oi] and [\sii] luminosities, respectively. In addition to the lower line fluxes estimated in this jet, we should also consider  that the blue-shifted jet is less collimated and is characterised by large bow shocks. Therefore the flux may be much more spread out and at larger distances the effect of sideways mass losses will be more pronounced than in the red-shifted lobe.

\begin{table*}
    \centering
    \begin{threeparttable}

        \caption[target]{Mass outflow rates and efficiencies in each jet lobe.}
        \renewcommand{\arraystretch}{1.2}
        \begin{tabular}{{ p{0.12 \textwidth}<{\raggedright} p{0.1\textwidth}<{\raggedright} p{0.13\textwidth}<{\raggedright} p{0.13\textwidth}<{\raggedright} p{0.13\textwidth}<{\raggedright} p{0.12\textwidth}<{\raggedright}}}
                \hline \hline
        & & \multicolumn{2}{l}{MUSE} & \multicolumn{2}{l}{X-shooter} \\
                 Jet lobe  & Method &  \makecell[l]{\Mout\ \\ (10$^{-9}$ \Msun~\peryr)} & \makecell[l]{Efficiency \\ (\Mout/\Macc)} & \makecell[l]{\Mout\ \\ (10$^{-9}$ \Msun~\peryr)} & \makecell[l]{Efficiency \\ (\Mout/\Macc)} \\
       Red-shifted  & & & &  &  \\
        & \nh\  & 9.2 \tiny{$\pm$} 3.9 & 0.038  & ... & ...   \\
         & L$_{[\oi]\lambda6300}$  & 8.9 \tiny{$\pm$} 4.6 & 0.037 & 7.43  \tiny{$\pm$}  3.83 & 0.03  \\ 
         &  L$_{[\sii]\lambda6731}$ & 5.6 \tiny{$\pm$} 2.4 & 0.023  & 5.65 \tiny{$\pm$} 2.14 & 0.02  \\
        Blue-shifted &     &  & & &  \\
                   &  \nh\  & 13.7  \tiny{$\pm$} 6.0 & 0.057 & ... & ...  \\
                   &  L$_{[\oi]\lambda6300}$ & 2.22 \tiny{$\pm$} 0.89 & 0.01 & ...  & ...  \\ 
         & L$_{[\sii]\lambda6731}$  & 0.66 \tiny{$\pm$} 0.19 & 0.003  & ... & ...   \\ 
                \hline
        \end{tabular} 
        \label{table:mass_outflow}
        \begin{tablenotes}
        \item[] \textbf{Notes:} Values are sampled at +2\arcsec\ and -2\arcsec\ from the source for the red- and blue-shifted shifted lobes, respectively, as this is the first position where the jet width is resolved in both lobes. The jet efficiency (\Mout/\Macc) is given assuming \av\ at the source of 2.5 mag.
        \end{tablenotes}
        \end{threeparttable}
   \end{table*}
    
    We expected the value of \Mout\ measured via the gas density, \nh\, to provide an upper estimate of the outflow rate. Indeed, in the red-shifted lobe this method produces the highest estimates of \Mout\ at almost every position. The outflow rate measured through \Ox\ luminosity is also higher than from the \siib\ luminosity. We can consider the ratio between the outflow rates from FEL luminosity and \nh\ to estimate the filling factor of the line-emitting gas. In the red-shifted lobe this is approximately constant along the jet in both species, with a mean value of 0.85 for [\oi] and 0.65 for [\sii] emission.
    
    Although the drop in density along the jet axis should theoretically be compensated by the increasing jet radius, we observed a small decreasing trend in the outflow rate along the jet. It is possible that this is due to the recombination of the gas reducing the FEL luminosity, but this is not consistent with the roughly constant \xe\ and \Te\ along the jet as shown by the results of the BE method. It is therefore more likely that the drop in \Mout\ is due to sideways losses of material through the internal working surfaces of the jet, with a possible contribution from time variability in the outflow.

    \section{Discussion}
    \label{section:discussion}  

    \subsection{Mass accretion rates and jet efficiencies}
    \label{subsection:accretion_discuss}
    
    We estimated an average  of \Macc\ = 2.4 $\times~10^{-7}$ \Msun~\peryr, assuming \av\ = 2.5 mag, or  \Macc\ = 5.53 $\times~10^{-8}$ \Msun~\peryr\ if we take \av\ = 1.26 mag. These values are consistent with typical accretion rates for CTTS sources, although the former value is approximately a factor of 4 higher than the 6.3 $\times~10^{-8}$ \Msun\ \peryr\ estimated by \citet{Comeron2010}. However, our estimates for \Macc\ derived from \caii\ $\lambda$8662, the accretion tracer used in their study, yields an average of 1.31 $\times~10^{-7}$ \Msun\ \peryr\ with \av\ = 2.5 mag and 4.95 $\times~10^{-8}$ \Msun\ \peryr\  with \av\ = 1.26 mag, indicating that our methodology yields results close to their findings. \citet{Alcala2017} previously made use of the same X-shooter dataset to derive log \Lacc\ = -2.0 from the UV excess in the Th 28 spectrum (although this study did not estimate the mass accretion rate). The UV excess is theoretically subject to obscuration, and indeed we find that before correction the permitted line tracers also yield average log \Lacc\ = -1.9 with \av\ = 1.5 mag, or -2.4 with \av\ = 1.26. 

    Our estimates then yield obscuration factors of  54 for \av\ = 2.5 and 44 for \av\ = 1.26. \citet{Nisini2018} also derived log \Lacc\ = 0.66 from the \Ox\ LVC emission of Th 28, which gives a similar obscuration factor of $\sim$ 59 when compared with the estimate from \citet{Alcala2017}. We can further compare this value with the case of ESO-\Ha\ 574, a jet with a similarly edge-on disc, but in which the accretion tracers are noted to be extremely suppressed \citep{Whelan2014b}. The obscuration factor estimated for the latter source is 150, suggesting that Th 28 is significantly less obscured in comparison. However, we note that although the [\oi] line is less obscured by the disc, there may still be some obscuration at this level and so our value may be underestimated. 
    
    It should additionally be noted that several accretion tracers included here show additional emission from the extended jet, such as the permitted H lines and \hei\ $\lambda$5876. Since the additional flux contribution cannot be separated from the accretion signatures this may inflate the estimated value of \Lacc. In order to test this, we sampled the line fluxes from the X-shooter data, using spectra centred at 1\farcs1 from the source along the axis of the bright red-shifted jet (the same methodology detailed in Sect. \ref{subsection:macc_estimates} was used to estimate the line fluxes).  We then compared our results with the values from the on-source sample. Emission lines with no detectable jet extensions (e.g., \heii\ 4686) yielded off-source fluxes approximately 16$\%$ that of the on-source value, with most lines showing less than 15$\%$  of the source flux at this distance. This residual flux is likely due to the seeing of 0\farcs9 compared with  the small offset and sampling area of 1\arcsec. As expected, the lines that remained relatively brightest in the off-source spectra were \caii\ K and H$\delta$, which both show extended emission along the jet axis. However, even in these cases, the fluxes in the off-source spectra were 25$\%$ and 22$\%$ that of the source spectra, respectively. These are not much higher proportions than found in the lines with no observed jet emission. We therefore conclude that the jet contributes only a small fraction of flux to the emission measured at the source, resulting in a slight over-estimation of \Macc.

    Given the uncertainties due to extinction, flux uncertainty, assumed stellar mass and distance, and accretion variability, our estimates agree well with the previously published measurements. We take note of this similarity since the \citet{Comeron2010} value is derived using estimates of the intrinsic properties of the source to compensate for the difference between the observed and intrinsic magnitudes due to obscuration and, hence, to obtain the total line flux, rather than using a comparison to the \Ox\ line. The consistency between results obtained using these alternate approaches suggests that our method provides a good estimate of the obscuration factor affecting Th 28, and that the [\oi] line is minimally subject to obscuration from the disc. The results from the combined MUSE and X-shooter datasets have the additional advantage that making use of a broad range of emission lines provides a more reliable estimate of \Macc.
    
    Taking the average value of \Macc\ to be 2.4 $\times~ 10^{-7}$ \Msun~\peryr, the jet efficiency (i.e., the ratio of mass outflow to accretion \Mout /\Macc) is 2-3 $\%$ in the red-shifted lobe. As discussed in Sect. \ref{subsection:macc_estimates}, this value can be obtained by assuming \av\ at the source of 2.5 mag. If we take \av\ = 1.26 mag (and, hence, \Macc\ = 5.53 $\times~ 10^{-8}$ \Msun~\peryr) we obtain efficiencies of 10-16 $\%$ in the red-shifted jet and  1-25 $\%$ in the blue-shifted jet. These values are in line with previously observed optical jet efficiencies of a few percent to $\sim$10$\%$, but underscore the impact of the assumed \av\ on our measurements.
  
    \subsection{Asymmetries among the jet lobes}
    \label{subsection:asymmetries}
 
        The asymmetries between the red- and blue-shifted jet lobes have been previously shown in terms of morphology and kinematics. Here, we have investigated the asymmetries in terms of extinction, physical conditions and mass outflow rates. Estimates using the \Ha/\Hb\ ratio (see Appendix \ref{section:extinction}) show that when taking into account the indications of higher shock velocities in the blue-shifted lobe, \av\ is slightly higher on this side; namely, by about 2-2.5 mag compared with 0-1.4 mag in the red-shifted lobe. This is consistent with the much brighter emission in the red-shifted lobe, although we note that $v_{\mathrm{shock}}$ is not well constrained in this region; a more detailed future analysis of the shock properties may help to clarify this. On the other hand, the difference in extinction is not large, and as discussed in Appendix \ref{section:extinction}, the values of \av\ in the blue-shifted lobe may be over-estimated due to the contribution of \Ha\ emission from bow shock wings along the line of sight. This suggests that the envelope is not significantly denser in the direction of the blue-shifted lobe, which is consistent with the findings of \citet{Melnikov2023} and would indicate that the asymmetries in jet properties originate very close to the source.

    In Sect. \ref{section:diagnostics}, we present both emission line ratio maps and results from the BE method analysis comparing the jet properties in each lobe. Both the MUSE and X-shooter data allow us to directly compare line ratios used in the BE method with ratios tracing higher-density regions of the jet close to the source. These establish the consistency in the jet asymmetry across multiple tracers and density regimes. The results in both the channel maps and PV maps indicate that the blue-shifted jet shows both higher \eden\ and higher excitation than the red-shifted jet. 

    By applying the BE method, we can then further quantify the asymmetry in \Te, \xe\,, and \nh\ and find that the blue-shifted jet is hotter, more ionised, and denser than the red-shifted jet. This is consistent with the trends from the line ratio maps, although the same caveats apply to the BE method results in the blue-shifted jet lobe as those given in Sect. \ref{subsection:mass_outflow}. Whereas \citet{Liu2014} estimated the blue-shifted jet to have an ionisation fraction that is two to three times higher than the red-shifted jet (and a lower hydrogen density), we instead find that \xe\ is only slightly higher in the blue-shifted jet and that this lobe is also more dense than the red-shifted jet.

    Comparing the mass outflow rates between the two jet lobes is complicated by the uncertainty in the blue-shifted jet density. Within the uncertainties, \Mout\ measured using the \nh\ density is the same in both jets; on the other hand, estimates using the emission line fluxes yield much lower outflow rates in the blue-shifted lobe even with the significantly higher jet velocity on this side. More reliable estimates of the jet flux and, hence, the density in this lobe are needed to make a full comparison. However, the overall picture could suggest a less collimated outflow leading to a much lower observed mass ejection efficiency. Complicating this assessment, there are the more complex kinematic structures in the less collimated lobe, with the true jet emission partially blended with at least some scattered light and potentially a strong LVC as well. To make a reliable determination of the true asymmetries between the two jet lobes, it will be necessary to account for these different components as well as the different morphology of the blue-shifted jet. A more detailed analysis of the shock properties and kinematics in this region is therefore necessary.

    \subsection{Asymmetry across the jet axis}
    \label{subsubsection:axial_asymmetry}
    
    The MUSE spectro-images of Th 28 show a small-scale precession of the inner jet axis. This signature could be caused by precession of the inner disc due to a companion of brown dwarf mass orbiting within 1 au of the jet source \citep{Murphy2021}. On the other hand, such offsets from the jet axis can also be caused by the presence of asymmetric shocks, as seen in HD 163296 \citep{Kirwan2022}. It is therefore a key point of interest to examine the transverse PV maps of the jet for signs of spatial asymmetry in the knots. In both jet lobes, the transverse maps show a very symmetric spatial structure in several knots, with a hot, ionised, and less dense core at higher velocities shifting to lower temperatures and higher densities at lower velocities and towards the spatial edges of the jet. The non-detection of transverse asymmetries in the extended jet suggests that the previously detected wiggling represents a real precession of the jet axis due to the influence of a companion object or warping of the inner disc \citep{Erkal2021}. Although the non-detection of an asymmetry near the source position is in contrast to the transverse asymmetry in electron density reported by Coffey et al 2008, that study sampled emission very close to the red-shifted jet base with higher spatial resolution (0\farcs3 from the source position); whereas, our study is seeing-limited (spatial resolution: 0\farcs9) and the emission near the jet base may include contamination from the blue-shifted lobe.
    
    \section{Conclusions}
    We present the results of a study of the asymmetric Th~28 jet using combined spectro-imaging and spectra from VLT/MUSE and X-shooter, respectively. We analyse key optical emission lines detected within the jet to estimate the gas properties and mass outflow rates within both lobes, and compare the latter with the estimated mass accretion rate onto the source. The main highlights of our results are summarised as follows:
    
    \begin{enumerate}
    
    \item{The X-shooter spectra show jet emission from a broad array of optical emission lines, including several which trace the elusive blue-shifted jet of Th 28. These spectra also show indications of a fast bow shock ($v_{\mathrm{shock}} >$ 200 \kms) between 1-2\arcsec\ from the source in the blue-shifted lobe. The [\oi] and [\sii] emission lines in this lobe also show a possible extended LVC component up to 2\arcsec\ from the source.}
    
    \item{We used optical emission line ratios and the BE method to map the density and excitation properties of the jet in both spectro-images and PV diagrams. Both methods highlight the asymmetric properties between the two jet lobes. The blue-shifted jet is estimated to be hotter than the inner red-shifted jet, with a moderately higher ionisation fraction. In contrast to previous studies, we also estimated substantially higher densities, \nh\,, in the blue-shifted lobe, almost twice what was obtained for its counterpart.}

    \item{In contrast, the transverse maps of these properties show strong axial symmetry across the jet axis in both lobes, with no significant indication of spatial asymmetries. This supports the previous finding of precession within the inner jet and suggests that the observed wiggling pattern is not caused by the effect of asymmetric shocks within the jet. }
    
    \item{We measured an average mass accretion rate of 2.4 $\times$ 10$^{-7}$ \Msun\ \peryr and estimated the obscuration factor due to grey scattering around the star to be $\sim$54. We find a substantial asymmetry in the mass outflow rates and, hence, efficiencies (\Mout /\Macc) between the two lobes. In particular, the red-shifted jet yields consistent efficiencies of 2-4$\%$, while the blue-shifted jet is much less consistent but yields efficiencies between $<~1\%$ and 6$\%,$ depending on the method used to estimate the mass outflow. Although \Mout\ is in itself a key observable parameter to constrain the disc magnetic field strengths in models of MHD disc wind launching, this asymmetry is a particularly significant finding. Recent simulations taking into account non-ideal MHD effects (particularly the Hall effect) have reproduced asymmetric disc winds with an enhancement of \Mout\ and outflow velocities on one side of the disc. The difference in mass outflow rate between both lobes ranges between a factor of a few and up to two orders of magnitude \citet{Bai2017, Bethune2017}. In our measurements, we find asymmetries in \Mout\ ranging from moderately higher in the blue-shifted lobe to an order of magnitude lower, due to the uncertainty associated with measurements on this side of the jet. Further works aimed at accurately constraining the mass outflow rate in this lobe will therefore be key to fully characterising the asymmetries in this jet and comparing them with models of asymmetric jet and wind launching.}
    
    \end{enumerate}

    Although the Th 28 jet poses a challenge in making optimal use of the combined capabilities of both MUSE and X-shooter, these results show that the capabilities of each instrument are invaluable in the study of complex regions of the jet close to the source in difficult targets. The spatial information provided by MUSE and the kinematic and spectral detail available with X-shooter make it possible to distinguish detailed structure across different regions of two very asymmetric jet lobes. The asymmetries in jet velocities, excitation and mass outflow rates, which originate close to the source and are likely intrinsic to the jet, make Th 28 a valuable case in constraining MHD models of asymmetric jet launching. A future study will make further use of these datasets to constrain the shock properties in each of the jet lobes and provide more insight into their asymmetries.

\section{Data availability}
\label{section:supplemental}
Supplementary materials for this paper are available via Zenodo at \url{https://doi.org/10.5281/zenodo.13373809}.

\begin{acknowledgements} 
We thank the referee for their comments which improved the quality of this paper. A. M. acknowledges the grant from the National Science and Technology Council (NSTC) of Taiwan 112-2112-M-001-031- and 113-2112-M-001-009-. A. M. and E. W. wish to acknowledge the support of the Irish Research Council under the Ulysses Programme. T. P. R. acknowledges support for ERC grant 743029 (EASY). JMA acknowledges financial support from PRIN-MUR 2022 20228JPA3A “The path to star and planet formation in the JWST era (PATH)” funded by NextGeneration EU and by INAF-GoG 2022 “NIR-dark Accretion Outbursts in Massive Young stellar objects (NAOMY)”. FB, BN, SA and JMA acknowledge financial support from the Large Grant INAF 2022 “YSOs Outflows, Disks and Accretion: towards a global framework for the evolution of planet forming systems (YODA)”.
\end{acknowledgements}  
\bibliographystyle{aa}
\bibliography{Bibliography}

        \clearpage

\begin{appendix}
   \FloatBarrier
   \section{Extinction along the jet lobes}

    \label{section:extinction}

    \begin{table*}
        \centering
        
        \caption[observations]{Estimated \av\ averaged over each region of the Th 28 jets}
        \renewcommand{\arraystretch}{1.8}
        \begin{tabular}{{p{0.15\textwidth}<{\raggedright} p{0.12\textwidth}<{\raggedright} p{0.12\textwidth}<{\raggedright} p{0.15\textwidth}<{\raggedright}  p{0.08\textwidth}<{\raggedright}}}
                \hline \hline
                Region &  Source offset &  Intrinsic \Ha/\Hb\ & Measured \Ha/\Hb\ & \av\ (mag)  \\
        \hline
                Blue-shifted lobe  & -5\arcsec\ to -1\farcs5 &  2.9 & 4.9 & 2.3 \\
        Source    &  -1 to +1  & 2.74 & 5.0 & 2.5 \\ 
        Red-Shifted lobe    & +1\farcs5 to 9\arcsec\ & 4.0 & 4.4 & 0.4 \\ 
                \hline
        \end{tabular} 
        \label{table:extinction}
        \end{table*}

    Previous estimates of extinction for Th 28 vary widely between 0.32 and 4.5 \citep{Hughes1994, Sartori2003, Evans2009, Mortier2011, Louvet2016}. However, \citet{Liu2014} show that assuming \av\ $>$ 3.5 mag at the source position would imply an unrealistically high X-ray luminosity for a CTTS, and most of these studies estimate the extinction between 2.5-2.9 mag. Since Th 28 is observed primarily in scattered light, this introduces an additional uncertainty as to whether the extinction at the source position can be estimated accurately. We therefore investigate the extinction in both jets and at the source using our observations.
        
\begin{figure}
        \centering
        \includegraphics[width=9cm, trim= 0cm 0.3cm 0cm 0cm, clip=true]{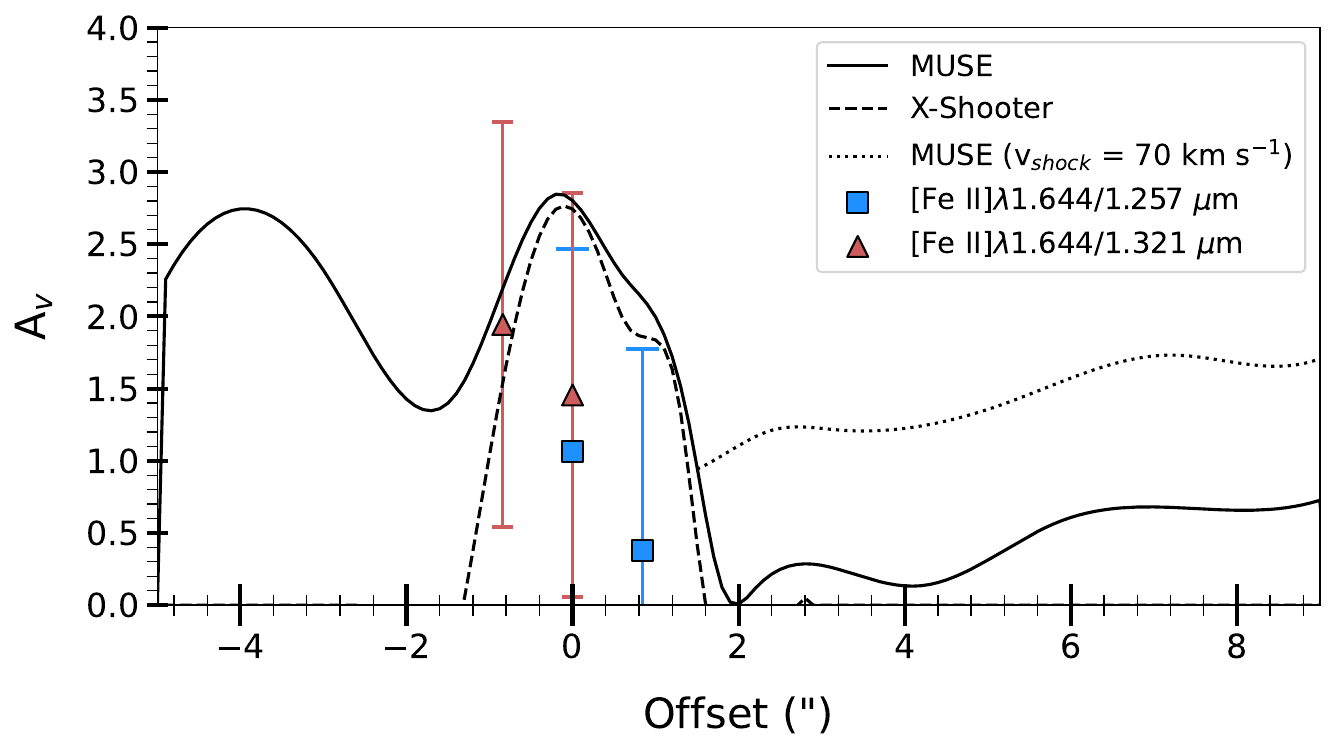}
        \caption{Comparison of the estimated extinction along the Th 28 jet lobes, using \Ha\ and \Hb\ fluxes sampled from the MUSE and X-shooter data and assuming typical $v_{\mathrm{shock}}$ of 40 \kms\ for the red-shifted lobe (right). The source position is marked at 0 and the dotted line shows the upper bound of \av\ values in the red-shifted jet assuming $v_{\mathrm{shock}}$ = 70 \kms.  The extinction values estimated using [\feii] fluxes from the X-shooter data are over-plotted.}
        \label{fig:Avfunc_comparison}     
\end{figure}

    We estimate the extinction along both jet lobes using the H Balmer $\alpha$ and $\beta$ emission lines. In the MUSE observations these are visible along both jet lobes as well as the western bow shock HHW. However, the wavelength range of the partial datacubes does not include H$\beta$; therefore, to ensure consistent flux calibrations we use the older datacube to sample both the H$\beta$ and H$\alpha$ line fluxes along the jet. The line spectra were sampled from 1\arcsec\ $\times$ 1\arcsec\ regions along the jet axis. We used the RedCorr() class of the Python package PyNeb \citep{Luridiana2015} which calculates the corresponding value of \av\ given the observed and theoretical ratios between two emission lines. We take the typical assumption for Lupus 3 of  R$_{v}$ = 5.5 \citep{Evans2009, Mortier2011} and use the extinction law described by \citet{Cardelli1989}. We thus obtain estimates of \av\ at each sampled position, and interpolate them with a polynomial to obtain a smooth extinction function along the jet (see Fig. \ref{fig:Avfunc_comparison}). The resulting function can then be applied to estimate the extinction correction at any position along the jet at any wavelength of interest. 
    
    We estimate the intrinsic Balmer decrement for each region of the jet using the models of \citet{Hartigan1994} to account for the effect of local shock speed on the relative ionisation. The presence of \OIII\ emission in the blue-shifted jet and in HHW indicates $v_{\mathrm{shock}}~ > 80-90$ \kms, and the expected decrement should therefore be close to 3.0 in these regions. The red-shifted jet has a significantly lower velocity and we assume lower $v_{\mathrm{shock}}~\sim 40$ \kms\ with an expected decrement $\sim$ 4.0. For the on-source position we assume a value of 2.86 for optically thick conditions \citep{Osterbrock1989}, given the high estimated densities and temperature $\sim$ 2 $\times 10^{4}$ K. However, we note that \citet{Antoniucci2017} have shown that the assumption of optically thick emission may not hold in T Tauri sources, so that the theoretical decrement is higher and consequently the true \av\ is lower. 

    Figure \ref{fig:Avfunc_comparison} shows the resulting profile of the estimated \av\ along the centre of the jet axis. The average extinction values for the main jet regions are summarised in Table \ref{table:extinction}. We find that the blue-shifted jet has relatively high extinction 1.5-2.8 mag; the red-shifted jet shows significantly less extinction at 0 - 0.6 mag, although some variation can be seen roughly corresponding to the known knot positions. Fig. \ref{fig:Avfunc_comparison} also shows the calculated extinction for the red-shifted jet assuming a maximum $v_{\mathrm{shock}}$ = 70 \kms, which yields an average \av\ = 1.4 mag. Taking the 5$\%$ flux calibration uncertainty estimated for for MUSE in each line this gives an approximate uncertainty of 7$\%$ for the ratio \bdec; by varying the input ratios to the RedCorr() routine by this value we estimate an uncertainty of 0.3 mag for the extinction values in all regions of the jet.  
    
    As expected, the highest extinction is found at the source position with a peak value of \av\ = 2.7 mag, in line with previous estimates. The blue-shifted jet shows higher extinction relative to the red-shifted lobe, which may partially account for the relative faintness of this lobe. On the other hand, the blue-shifted jet emission may include unresolved bow shocks as are typical of the structures in this lobe. The sampled emission therefore likely includes emission from the wings where the effective shock velocity is lower than we have assumed for this region, and hence \Ha\ emission is enhanced relative to \Hb, increasing the observed extinction \citep{Hartigan1987, Hartigan1994}. The extinction in this lobe may therefore be over-estimated in our results.
    
    As a comparison, the same procedure was carried out using fluxes sampled from the X-shooter observations. Due to the shorter exposure time, the \Hb\ line was not visible in much of the blue-shifted lobe, however both emission lines can be sampled near the source position and in the inner red-shifted jet. We also made use of the [\feii]$\lambda$ 1.644/1.257 $\mu$m and 1.644/1.321 $\mu$m line ratios using spectra sampled from the NIR arm. We used intrinsic values for these ratios of 0.88 and 3.16, respectively, using the values derived empirically by \citet{Giannini2015}. The measured ratios fall below the intrinsic values at all positions more than 0\farcs4 from the source position; at the source we find estimated \av\ of $\sim$1 mag, lower than that estimated from the H lines. The resulting \av\ estimates are also shown in Fig. \ref{fig:Avfunc_comparison} and are overall lower than the estimates found with the MUSE data with an average \av\ = 1.26 at the source position. To maximise the consistency between different jet regions, the MUSE results are used for all extinction corrections to flux values in the rest of our results -- except where otherwise specified (see comparison of mass accretion rates using different \av\ values in Sect. \ref{subsection:macc_estimates}).          

        \end{appendix}

\end{document}